\documentclass[11pt]{article}
\usepackage[left=2.5cm,right=2.5cm,top=2cm,bottom=3cm]{geometry}
\usepackage{amsmath,amssymb,graphicx,color,array}

\numberwithin{equation}{section}
\makeatletter

\@addtoreset{equation}{section}
\makeatother

\graphicspath{{eps/}}

\newcommand{\vac}{(\ \ )}
\newcommand{\dcup}{(\begin{smallmatrix} \Cup \\ \phantom{\Cap} \end{smallmatrix})}
\newcommand{\dcap}{(\begin{smallmatrix} \phantom{\Cup} \\ \Cap \end{smallmatrix})}
\newcommand{\idl}{(\rangle\!\rangle\: \phantom{\langle\!\langle})}
\newcommand{\idr}{(\phantom{\rangle\!\rangle\: } \langle\!\langle)}
\newcommand{\idlr}{(\rangle\!\rangle\: \langle\!\langle)}
\newcommand{\vup}{(\diagup\!\!\!\!\diagup)}
\newcommand{\vdown}{(\diagdown\!\!\!\!\diagdown)}

\newcommand{\resp}{resp.}
\newcommand{\On}{{\rm O}(n)}
\newcommand{\Rc}{\check{R}}
\newcommand{\SU}{{\rm SU}}
\newcommand{\Sp}{{\rm Sp}}
\newcommand{\up}{\uparrow}
\newcommand{\dow}{\downarrow}
\newcommand{\half}{\frac{1}{2}}
\newcommand{\halfpi}{\frac{\pi}{2}}

\newcommand{\idop}{{\bf 1}}

\newcommand{\wt}{\widetilde}
\newcommand{\nodag}{{\phantom{\dag}}}

\title{An integrable modification of the critical Chalker-Coddington network model}

\author{
  Yacine Ikhlef
  \smallskip \\
  Section Math\'ematiques, Universit\'e de Gen\`eve \\
  2-4 rue du Li\`evre, CP 64, 1211 Gen\`eve 4, Switzerland
  \bigskip \\
  Paul Fendley
  \smallskip \\
  Microsoft Research, Station Q, University of California \\
  Santa Barbara, CA 93106 \\
  {and} \\
  Department of Physics, University of Virginia \\
  Charlottesville, VA 22904-4714, USA
  \bigskip \\
  John Cardy
  \smallskip \\
  Rudolf Peierls Centre for Theoretical Physics, University of Oxford \\
  1 Keble Road, OX1 3NP, United Kingdom \\
  and All Souls College, Oxford
}

\begin{document}

\maketitle

\begin{abstract}
  We consider the Chalker-Coddington network model for 
  the Integer Quantum Hall Effect, and examine the possibility of
  solving it exactly. In the supersymmetric path integral
  framework, we introduce a truncation procedure, leading to a series
  of well-defined two-dimensional loop models, with two loop flavours.
  In the phase diagram of the first-order truncated model, we identify
  four integrable branches related to the dilute Birman-Wenzl-Murakami
  braid-monoid algebra, and parameterised by the loop fugacity $n$.
  In the continuum limit, two of these branches (1,2) are described by a pair of
  decoupled copies of a Coulomb-Gas theory, whereas the other two branches (3,4)
  couple the two loop flavours, and relate to an $\SU(2)_r \times \SU(2)_r / \SU(2)_{2r}$
  Wess-Zumino-Witten (WZW) coset model for the particular values $n= -2\cos[\pi/(r+2)]$
  where $r$ is a positive integer. The truncated Chalker-Coddington model is the $n=0$
  point of branch 4. By numerical diagonalisation, we find that its universality
  class is neither an analytic continuation of the WZW coset, nor the
  universality class of the original Chalker-Coddington model. It constitutes
  rather an integrable, critical approximation to the latter.
\end{abstract}

\section{Introduction}
\label{sec:intro}

The transition between plateaux in the Integer Quantum Hall Effect
(IQHE) is a quantum critical phenomenon, which was predicted
theoretically~\cite{LLP,Pruisken} and observed
experimentally~\cite{QHE-exp} a few decades ago. Although
experimentally there is no a priori reason to neglect
electron-electron interactions, it is usually modelled
theoretically by noninteracting particles in two dimensions (2d), in a
perpendicular magnetic field and a random potential. Despite the
apparent simplicity of this conceptual setup, it turns out to  be
very difficult to derive analytically the critical exponents of
this transition. Important progress was achieved by the
introduction of a simple network model which retains the salient
features of guiding centre motion and quantum tunnelling in the
presence of disorder: the Chalker-Coddington (CC) model~\cite{CC}.
Extensive numerical studies based on the CC model
or other approaches have led to good estimates for the critical
exponents, notably the correlation-length exponent $\nu = 2.37 \pm 0.02$~\cite{Cain}
(a larger value $\nu=2.593\pm 0.006$ has also been reported~\cite{Slevin}).
Also, a semi-classical argument~\cite{QHE-semi} yields the prediction $\nu=7/3$.

The CC model is also the starting point for several analytical approaches, like
the description by a $\sigma$-model~\cite{Ludwig}, or
a mapping to a one-dimensional (1d) quantum many-body system~\cite{KM,MT}.
However, from the point of view of critical lattice models, no
exact solution of the CC model has been found so far.

The situation is very different for the {\it spin} Quantum Hall
Effect (SQHE): the generalisation of the CC model to
SQHE~\cite{CC-SU2} (which we shall call $\Sp(2)$-CC) maps exactly
to classical bond percolation, where a large class of exponents
are known~\cite{DS}. This mapping of $\Sp(2)$-CC to classical
percolation was first observed by Gruzberg {\it et al.}~\cite{Gruzberg}, who used
a supersymmetric (SUSY) spin-chain formulation. Later on, it was
realised~\cite{CC-path1,CC-path2} that the SUSY {\it lattice path integral}
maps $\Sp(2)$-CC to a statistical model of lattice paths, which
are exactly the hulls of bond-percolation clusters. Moreover, a
number of SQHE physical observables are expressed in terms of
percolation correlation functions, and this mapping is valid even
at the level of lattice models.

In this paper, we propose a treatment of the original CC model based 
on the lattice path integral. Since the corresponding statistical model 
involves paths which may pass through a given edge
infinitely many times, the number of configurations per unit surface
is infinite, and the model is not directly tractable by exact-solution 
methods such as Yang-Baxter integrability and Conformal Field Theory (CFT).
We therefore introduce a {\it truncation procedure}, leading to a series of
finite statistical models, and focus on the first order of truncation.
The arising model is a two-colour loop model including vacancies, and with loop
fugacity $n=0$.

Integrable multi-colour loop models have been known for a long
time~\cite{multi-colour}. They were originally defined through
multi-dimensional height models, but they may as well describe
coupled copies of classical magnetism models, such as the Potts or
$\On$ models, and also the ground state of {\it quantum} loop
models. More specifically, in a two-colour, completely
packed ({\it i.e.} without vacancies) loop
model~\cite{Martins-Nienhuis,Paul-Jesper}, new integrable points
were identified through a mapping to a braid-monoid algebra: the Birman-Wenzl-Murakami
(BWM) algebra~\cite{BWM}. In the present paper, we use a similar approach on
the loop model arising from our truncation procedure, which is a
two-colour loop model including vacancies. Generalising to
arbitrary loop fugacity $n$, we obtain four critical branches in
the phase diagram of this loop model. We then study the critical
properties of these branches.

We find that two of these regimes (denoted 1 and 2) correspond to a pair
of decoupled Coulomb-Gas (CG) theories, whereas the other two (3 and 4)
relate to the $\SU(2)_r \times \SU(2)_r / \SU(2)_{2r}$ Wess-Zumino-Witten
coset model, for values $n= \pm 2 \cos \frac{\pi}{r+2}$ with $r\in\{1,2,3,\dots\}$.
We obtain analytically two critical exponents: one of them, $X_{\rm int}$, corresponds
to an elliptic deformation of the integrable weights, and the other one,
$X_{(1,1;{\rm adj})}$, is associated to a perturbation of the weight per monomer.
The truncated, modified CC model is realised by the $n=0$
point of regime 4, but this point is outside the validity range
for the analytic continuation of the WZW exponents.
Our numerical study gives the estimate $\nu \simeq 1.1$ for the correlation-length
exponent, and $d_f \simeq 1.71$ for the fractal dimension of paths.
This is clearly incompatible with the IQHE universality class, and hence our
integrable two-colour loop model is only a crude approximation to IQHE.
However, the truncation procedure may be carried out to higher orders,
possibly yielding more accurate, solvable approximations.

The plan of the paper is as follows. In Section 2, we recall
the definition of the CC model and its lattice SUSY path-integral formulation,
and explain our truncation procedure,resulting in a two-colour
loop model. This truncation is compared in detail with the one used in~\cite{KM,MT}.
In Section 3, we use a mapping to a dilute braid-monoid algebra to derive
the integrable Boltzmann weights of the two-colour loop model,
as well as the corresponding 1d Hamiltonian.
In Section 4, we identify the four critical regimes of the integrable 
model and the corresponding CFTs. Numerical and analytical support for 
the identification of these CFTs is given.
In Section 5, we examine in more detail regime 4, which contains the truncated, 
modified CC model at $n=0$. We discuss the analytic continuation of CFT results,
and estimate numerically some critical exponents, including the correlation-length
exponent $\nu$.

The paper has three appendices. Appendix A contains the details of the mapping
to the dilute BWM (dBWM) algebra used in Section 3. In Appendix B, we exhibit a lattice 
holomorphic parafermion $\psi_s(z)$ in the integrable model. In Appendix C, we expose
the exact solution of a particular point in regime 4, which is mapped to free fermions.
This mapping provides a valuable check on our results, and also gives a proof
that the ${\rm O}(n=1)$ loop model has central charge $c=\half$.

\section{Truncation of the Chalker-Coddington model}
\label{sec:CC}

\subsection{The Chalker-Coddington model}
\label{sec:intro-CC}

The Chalker-Coddington model~\cite{CC} is a simple lattice model for
the IQHE. The latter consists of a two-dimensional gas of non-interacting 
electrons in a disordered medium, subject to a strong transverse magnetic
field. In the presence of the random potential, the Landau levels are
broadened, and eigenenergies are of the form $E=\left( k+\half \right) \hbar \omega_c + V_0$,
where $k$ is an integer, $\omega_c$ is the cyclotron energy
of the electron in the magnetic field, and $V_0$ is a random part.
Let us recall briefly the main ingredients of the CC model.

\begin{figure}[th]
  \begin{center}
    \begin{tabular}{m{4cm}m{4cm}}
      \includegraphics{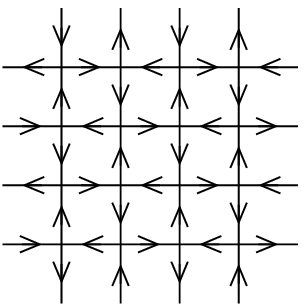}
      & \includegraphics{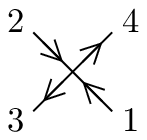} \\
      (a) & (b)
    \end{tabular}
    \caption{
      (a) Oriented square lattice $\cal L$ for the Chalker-Coddington model.
      (b) Labelling of the edges adjacent to a vertex of $\cal L$.
    }
    \label{fig:cc-def}
  \end{center}
\end{figure}

We consider an electron in the
eigenstate of energy $E$. The spatial trajectories
of the electron over finite time steps $\Delta t$ are modelled by paths on the
directed square lattice $\cal L$
(see Fig.~\ref{fig:cc-def}),
and the time-evolution operator over $\Delta t$ is denoted $\cal U$.
The operator $\cal U$ reads
\begin{equation}
  {\cal U} = \bigotimes_{{\rm edge} \ e} U_e
  \ \bigotimes_{{\rm vertex} \ v} U_v \,,
\end{equation}
with two types of factors:
\begin{itemize}
\item On each directed edge $e$, the operator $U_e$
takes the particle along $e$ and multiplies the wavefunction by a
random Aharonov-Bohm phase $\exp(i\phi_e)$, where the $\phi_e$ are
independent and uniformly distributed on the interval $[0,2\pi]$.
\item At each vertex $v$, the operator $U_v$ scatters the particle to one
of the outgoing edges.
In the bases $(1,2)$ and $(3,4)$ of Fig.~\ref{fig:cc-def}, $U_v$ is represented by the
unitary matrix:
\begin{equation}\label{eq:Smatrix}
  S = \left( \begin{array}{cc}
      \tanh \beta & 1/{\cosh \beta} \\
      1/{\cosh \beta} & -\tanh \beta
    \end{array} \right) \,.
\end{equation}
\end{itemize}
The parameter $\beta$ measures the distance to the plateau transition at
$E=E_c=\left(k+\half\right)\hbar \omega_c$. The critical value is
$\beta_c=\log(1+\sqrt{2})$, and the corresponding energy perturbation is assumed to behave
as~\cite{CC}
\begin{equation}
  (E-E_c) \propto (\beta-\beta_c) \,.
\end{equation}

No exact solution of the CC model is known, in the sense that the critical
exponents have not been determined analytically. However, very
good numerical estimates exist for some of these
exponents~\cite{Huck,Cain,Slevin}. In particular, the correlation-length
exponent $\nu_{\rm CC}$, defined by the scaling of the correlation
length
\begin{equation} \label{eq:nu-CC}
  \xi \propto |E-E_c|^{-\nu_{\rm CC}} \,,
\end{equation}
has been estimated as~\cite{Cain}
\begin{equation}
  \nu_{\rm CC} \simeq 2.37 \pm 0.02 \,.
\end{equation}

\subsection{Path integral representation}
\label{sec:path-int}

The problem of solving the CC model amounts to the diagonalisation
of a random time-evolution operator. We want to perform the
average over disorder, in order to turn this into a
translationally invariant 2d classical model. For
this purpose, we use the supersymmetric path integral
representation~\cite{Efetov}. The following derivation is very
analogous to what was done by one of us for the SQHE~\cite{CC-path2},
and we use the notations of~\cite{CC-path2} throughout this Section.

The Green's function between two edges $e_1$ and $e_2$ is:
\begin{equation}
  G(e_2,e_1,z) := \langle e_2 | (1-z {\cal U})^{-1} | e_1 \rangle \,.
\end{equation}
Here $z$ is a parameter which plays the role of the energy in the
usual Green's function $(E-{\cal H})^{-1}$: roughly speaking
$z\sim e^{{\rm i}E}$, where $E$ is measured from the filled Landau
level. We label $e_L, e_R$ the ends of any edge $e$, with the
convention that it is directed in the sense $e_R \to e_L$, and we
introduce the complex variables $b_L(e), b_R(e)$. The Gaussian
measure is defined as:
\begin{equation}
  \int [{\rm d} b] \ (\dots) :=
  \frac{1}{\pi} \int {\rm d}({\rm Re} \ b) \ {\rm d}({\rm Im} \ b)
  \ \exp(-b^*b) (\dots) \,,
  \qquad [{\rm D}b]:=\prod_e [{\rm d}b_L(e)] [{\rm d}b_R(e)] \,.
\end{equation}
The Green's function can then be written as a Gaussian integral on the $b_L(e), b_R(e)$:
\begin{equation} \label{eq:G-boson}
  G(e_2,e_1,z) = \frac{\int [{\rm D}b] \ b_L(e_2) b_L^*(e_1) \exp A_b}
  {\int [{\rm D}b] \ \exp A_b} \,,
\end{equation}
where the action reads
\begin{eqnarray}
  A_b &=& A_b^{({\rm edge})} + A_b^{({\rm vertex})} \,, \\
  A_b^{({\rm edge})} &=& z \sum_{{\rm edge} \ e}
  b^*_L(e) \exp(i \phi_e) b_R(e) \,, \label{eq:edge}\\
  A_b^{({\rm vertex})} &=& \sum_{{\rm vertex} \ v}
  \ \sum_{\begin{smallmatrix} i &\to&j \\ &v& \end{smallmatrix}}
  b^*_R(e_i) S_{ij} b_L(e_j) \,,
\end{eqnarray}
and the notation $\begin{smallmatrix} & \phantom{v} & \\ i &\to&j \\ &v& \end{smallmatrix}$ means
that $i$ ({\resp} $j$) is an incoming
({\resp} outgoing) edge adjacent to $v$.
The next step is to express the denominator in~\eqref{eq:G-boson} as the inverse of
a Gaussian integral over Grassmann variables $f_{L,R}(e), \bar{f}_{L,R}(e)$:
\begin{equation} \label{eq:G-SUSY}
  G(e_2,e_1,z) = \int [{\rm D}b] [{\rm D}f] \ b_L(e_2) b_L^*(e_1) \exp (A_b + A_f) \,,
\end{equation}
with the measure
\begin{equation}
  \int [{\rm d} f] \ (\dots) :=
  \int {\rm d}\bar{f} \ {\rm d}f \ \exp(-\bar{f} f) (\dots) \,,
  \qquad {[{\rm D}f]} := \prod_e [{\rm d}f_L(e)][{\rm d}f_R(e)] \,,
\end{equation}
and $A_f$ is the analog of $A_b$, with $b,b^*$ replaced by $f,\bar{f}$.

We denote by an overbar the quenched average over the variables $\phi_e$. A useful formula
for this computation is
\begin{equation} \label{eq:quench}
  \frac{1}{2\pi} \int_0^{2\pi} {\rm d}\phi \exp \left(
    u e^{i\phi} + v^* e^{-i\phi}
  \right) = \sum_{m=0}^\infty \frac{(u v^*)^m}{(m!)^2} \,.
\end{equation}
It easy to see, for instance, that $\overline{G(e_2,e_1,z)} = \delta(e_1,e_2)$.
When studying transport properties, the main quantity of interest is $\overline{|G|^2}$.
We write
\begin{eqnarray}
  |G(e_2,e_1,z)|^2 &=&
  \int [{\rm D}b] [{\rm D}f] \ b_L(e_2) b_L^*(e_1) \ e^{A_b + A_f}
  \times \int [{\rm D}b] [{\rm D}f] \ b_L^*(e_2) b_L(e_1) \ e^{A_b^* + A_f^*} \nonumber \\
  &=& \int [{\rm D}b_{1,2}] [{\rm D}f_{1,2}]
  \ b_{L1}(e_2) b_{L1}^*(e_1) b_{L2}^*(e_2) b_{L2}(e_1)
  \ e^{A_{b1} + A_{f1} + A_{b2}^* + A_{f2}^*} \,.
\end{eqnarray}
Using~\eqref{eq:quench}, we get:
\begin{eqnarray}
  && \overline{|G(e_2,e_1,z)|^2} =
  \int [{\rm D}b_{1,2}] [{\rm D}f_{1,2}]
  \ b_{L1}(e_2) b_{L1}^*(e_1) b_{L2}^*(e_2) b_{L2}(e_1) \times \nonumber \\
  && \quad \exp\left[ A^{({\rm vertex})}_{b1} + A^{({\rm vertex})*}_{b2}
    + A^{({\rm vertex})}_{f1} + A^{({\rm vertex})*}_{f2} \right] \times \nonumber \\
  && \quad \prod_e \sum_{m_e=0}^{\infty} \frac{(z^*z)^{m_e}}{(m_e!)^2}
    \left\{
      \left[ b_{L1}^*(e) b_{R1}(e) + \bar{f}_{L1}(e) f_{R1}(e)
      \right]
      \left[ b_{R2}^*(e) b_{L2}(e) + \bar{f}_{R2}(e) f_{L2}(e)
      \right]
    \right\}^{m_e} \,.
  \label{eq:G2}
\end{eqnarray}

The expression~\eqref{eq:G2} for $\overline{|G|^2}$ can be
interpreted graphically as follows. Each term in the expansion of
the product corresponds to a pair of paths $(\gamma_1,\gamma_2)$,
where $\gamma_1$ respects the orientation of the lattice $\cal L$
(forward path) and $\gamma_2$ follows the reverse orientation
(backward path). The two paths must use each edge $e$ the same
number of times $m_e$. Paths configurations are weighted by the
elements of the vertex $S$-matrix, and an additional factor
$(z^*z)^{m_e}$. Note that closed loops have a vanishing weight,
because the bosonic and fermionic contributions cancel each other.

\subsection{Truncation procedure}
\label{sec:truncation}

In the form~\eqref{eq:G2}, $\overline{|G|^2}$ can be viewed as a
two-point correlation function in a classical, two-dimensional
statistical model for two-colour path configurations. No
approximation has been introduced so far, and thus \eqref{eq:G2}
is identical to the value of $\overline{|G|^2}$ in the original CC
model. The main difficulty in evaluating \eqref{eq:G2} is that the
paths $\gamma_1, \gamma_2$ may go through a given edge an
arbitrary number of times $m_e$, and thus the statistical model
has an infinite number of degrees of freedom per edge. This type
of problem is not usually tractable by exact solution methods, so
we need to {\it truncate} the statistical model to a finite loop
model in order to use these methods. This is very analogous to
what Nienhuis did for the $\On$ spin model~\cite{nienhuis-On} on
the hexagonal lattice: in that context, the spin model with
variables $( {\bf S}_j \in \mathbb{R}^n, {\bf S}_j^2=1 )$ was
formally mapped to a polygon model where edges could be used an
arbitrary number of times, but the substitution $e^{J {\bf S}_i
\cdot {\bf S}_j} \to 1 + J  {\bf S}_i \cdot {\bf S}_j$ in the edge
interaction led to a finite loop model, while preserving the $\On$
symmetry of the original spin model.

\begin{figure}[th]
  \begin{center}
    \includegraphics{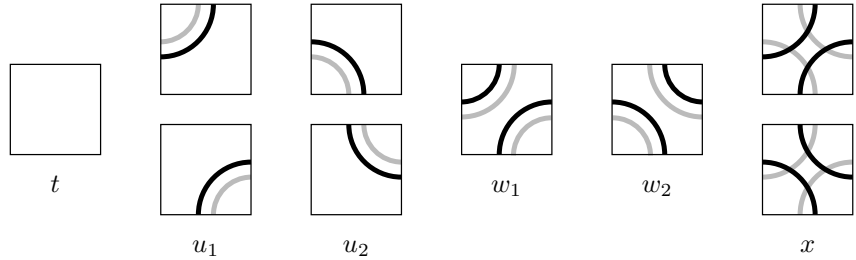}
    \caption{Vertices of the loop model arising from the truncation of the CC model.}
    \label{fig:cc-model}
  \end{center}
\end{figure}

The truncation we propose consists in keeping only the terms
of~\eqref{eq:G2} with $m_e \in \{0,1\}$, {\it i.e.} the
configurations where each of the paths $\gamma_1,\gamma_2$ visits
an edge at most once. This preserves the boson/fermion
supersymmetry, ensuring that closed loops still have a vanishing
weight in the truncated model. This can be seen as follows. In the
original expression~\eqref{eq:edge} for the action on the edges,
we can imagine choosing a different fugacity $z_e$ for each edge
(so that it now appears inside the summation over $e$.) This does
not affect the supersymmetry of the action. On expanding in powers
of all the $z_e$, the bosonic contribution to a given edge now
enters with a factor $(z_e^* z_e^{\phantom{*}})^{m_e}$. Thus our truncation to
$m_e\in\{0,1\}$ amounts to keeping only the terms up to first
order in the expansion of the partition function in powers of
$z_e^* z_e^{\phantom{*}}$, and then setting all the $z_e=z$ again.. Note that to
this order we have either nothing, or a pair of bosons of
different flavours (1 and 2), or a pair of fermions of different
flavours, propagating along each edge. The supersymmetry ensures
that each closed loop is counted with weight $0$. At this stage it
is simpler to switch to a replica formulation rather than using
supersymmetry explicitly: we have a model with two flavours of
boson, such that each edge is either unoccupied, or occupied by
each flavour exactly once. Each closed loop is counted with a
fugacity $n$, taking then $n=0$. The vertices are shown in
Fig.~\ref{fig:cc-model}.

We briefly comment on how higher order truncations would look in
this expansion. For example, at ${\rm O}\big((z_e^* z_e^{\phantom{*}})^2\big)$ we would
have either 2 pairs of bosons of each flavour, or 1 pair of bosons
and 1 pair of fermions. (We can never have more than one pair of
fermions because the Grassmann variables square to zero.) Note
that in such a truncation we could give such a configuration a
weight different from $(z_e^* z_e^{\phantom{*}})^2$ and still preserve the
supersymmetry. This points to the existence of an
infinite-dimensional space of possible supersymmetric truncations.
However in this paper we consider only the simplest.

We denote by $\overline{|G(e_2,e_1,z)|^2_{\rm tr}}$ the truncated analog of
$\overline{|G(e_2,e_1,z)|^2}$: $\overline{|G(e_2,e_1,z)|^2_{\rm tr}}$ is given by
the same expression as~\eqref{eq:G2}, but with the sum running
only over $m_e=0,1$. Then $\overline{|G(e_2,e_1,z)|^2_{\rm tr}}$ is interpreted as
a two-point function in the loop model defined by the loop vertices
of Fig.~\ref{fig:cc-model} and with loop weight $n=0$.

In the original CC model, the parameter~$\beta$ in the $S$-matrix
(\ref{eq:Smatrix}) is staggered. It is useful to consider an
anisotropic version of this, where it takes the value $\beta$ on
the even sublattice of $\cal L$ and $\beta'$ on the odd
sublattice. In this anisotropic CC model, the critical line
is~\cite{CC}:
\begin{equation} \label{eq:critical-beta}
  \sinh \beta \ \sinh \beta' = 1 \,.
\end{equation}
The Boltzmann weights of the truncated loop model are defined in
Fig.~~\ref{fig:cc-model}.   For general $\beta,\beta'$ they take
the values:
\begin{equation} \label{eq:weights-cc1}
  \begin{array}{lcl}
  t\phantom{'}, u_1, u_2, w_1, w_2, x\phantom{'}
  = 1, \ a\phantom{'}, \ b\phantom{'},
  \ a^2\phantom{'}, \ b^2\phantom{'}, \ -a \phantom{'} b
  &\quad& \hbox{(even sublattice),} \\
  t', u'_1, u'_2, w'_1, w'_2, x' = 1, \ b', \ a', \ b'^2, \ a'^2, \ -a'b'
  &\quad& \hbox{(odd sublattice),}
  \end{array}
\end{equation}
where
\begin{equation} \label{eq:weights-cc2}
  \begin{array}{lcl}
    a\phantom{'}:=  z^2 \cosh^{-2} \beta\phantom{'} \,,
    &\quad& b\phantom{'} := z^2 \tanh^2 \beta\phantom{'}  \,, \\
    a':= z^2 \cosh^{-2} \beta' \,,
    &\quad& b':= z^2 \tanh^2 \beta' \,.
  \end{array}
\end{equation}
Note that at the isotropic point these weights are
\begin{equation} \label{eq:isocritical}
  1, \ z^2/2, \ z^2/2, \ z^4/4, \ z^4/4, \ -z^4/4 \,.
\end{equation}

\subsection{Critical properties}
\label{sec:critical-prop}

We now discuss the observables of the model. The most important is
the mean square Green's function between two edges
$\overline{|G(e_2,e_1,z)|^2}$. At $z=1$ in the untruncated model
this gives the point conductance. For a system without any open
boundary contacts, this is identically equal to one by
conservation of probability. It is given by the sum over all pairs
of Feynman paths going out and back from $e_1$ to $e_2$, such that
each edge is traversed the same number of times in the forward
path as in the return path, and weighted by the appropriate
$S$-matrix elements of the CC model. It has been argued
\cite{gruzberg-unpub} that the weights for such `pictures' are all
positive. For $z \to 1^{-}$ and on the critical line
\eqref{eq:critical-beta} we expect a scaling form
\begin{equation} \label{eq:G2scaling}
  \overline{|G(e_2,e_1,z)|^2} \sim
  r^{-2X_{|G|^2}} \, F \left[ r(1-z)^{1/d_f} \right] \,,
\end{equation}
where $r=|e_1-e_2|$, $X_{|G|^2}=0$, and $d_f$ is the fractal dimension of
these pictures (whereby their total mass $M$ behaves as
$r^{d_f}$). The absence of the prefactor $r^{-2X_{|G|^2}}$ is a
consequence of the fact that $|G(z=1)|^2=1$ for a closed system.

\begin{figure}[th]
  \begin{center}
    \includegraphics{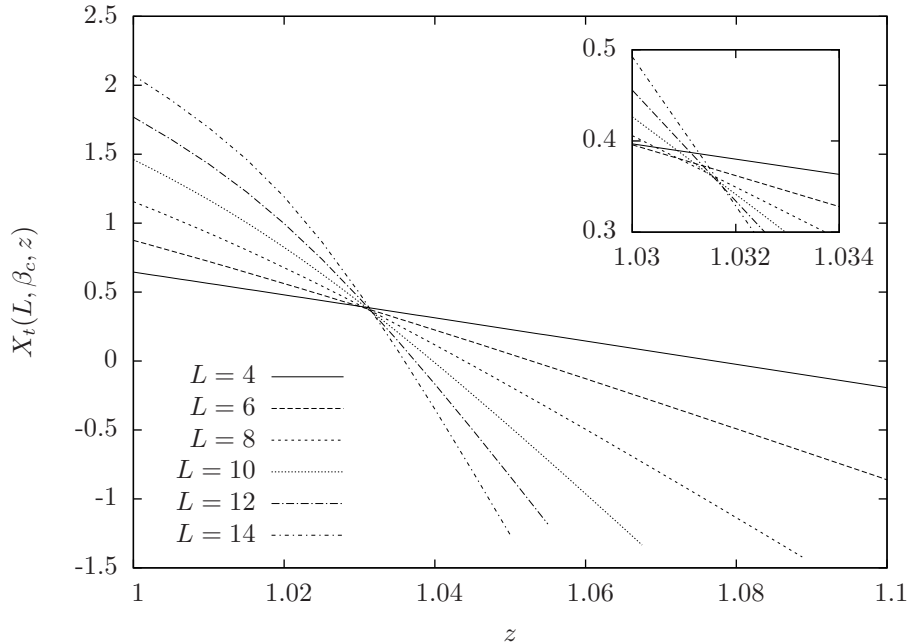}
    \caption{Numerical determination of the critical monomer fugacity $z_c$ in
      the model of Fig.~\ref{fig:cc-model}. On the $y$-axis is plotted the
      effective thermal exponent $X_t(L,\beta_c,z) = \frac{L}{2\pi} \log \frac{\Lambda_0}{\Lambda_1}$.
    }
    \label{fig:zc-trcc}
  \end{center}
\end{figure}

In the truncated model, we no longer have probability conservation
and so the point $z=1$ is no longer special. Instead, in analogy
with other loop models, we expect to find a different critical
point, at $z=z_c$, say, such that the average loop length is
finite for $z<z_c$ and diverges for $z\geq z_c$. The point
conductance $\overline{|G(e_2,e_1,z)|^2_{\rm tr}}$ now corresponds to the
weighted sum of a pair of black and grey paths connecting $e_1$
and $e_2$. On the critical line and as $z\to z_c^{-}$ we expect the
same scaling form as in \eqref{eq:G2scaling} with $(1-z)$ replaced by $(z_c-z)$,
but not necessarily with same exponents as in the full model.
In Figure~\ref{fig:zc-trcc} and Table~\ref{table:zc}, we show the numerical determination
of $z_c$ using the two largest eigenvalues $\Lambda_0,\Lambda_1$ of the transfer matrix.
These eigenvalues define the thermal exponent $X_t$ through the CFT form
of the free-energy gap:
\begin{equation}
  \log \frac{\Lambda_0}{\Lambda_1} \simeq \frac{2\pi X_t}{L} \,.
\end{equation}

\begin{table}
  \begin{center}
    \begin{tabular}{c|r|r|r|r}
      $L$ & 4 & 6 & 8 & 10 \\
      \hline
      $z_c(L,L+2)$ & 1.029885 & 1.030895 & 1.031454 & 1.031695
    \end{tabular}
    \caption{Finite-size estimates of the critical monomer fugacity $z_c$ in the
      model of Fig.~\ref{fig:cc-model}. The value $z_c(L,L+2)$ is defined as the solution of
      $X_t(L,\beta_c,z)=X_t(L+2,\beta_c,z)$, where $X_t(L,\beta,z)= \frac{L}{2\pi} \log \frac{\Lambda_0}{\Lambda_1}$ is the
      effective thermal exponent.}
    \label{table:zc}
  \end{center}
\end{table}

More generally, it is possible to consider `watermelon' exponents
$X_{\ell_1,\ell_2}$ corresponding to $\ell_1$ black and $\ell_2$
grey paths originating from the vicinity of a given edge. The
truncation constraint of course implies that these cannot
originate on the same edge for $\ell>1$, but we imagine taking the
scaling limit where edges a finite distance apart on the lattice
are mapped to the same point. These operators are well suited for
a transfer-matrix-based numerical analysis~\cite{ds-tm}.

In particular we see that $X_{1,1}$ corresponds to $X_{|G|^2}$ in~\eqref{eq:G2scaling}.
Also, since $z^*z$ counts the number of edges connected to 2 black and 2 grey paths,
we have
\begin{equation}\label{eq:X22}
  d_f = 2-X_{2,2}\,.
\end{equation}
Using transfer-matrix diagonalisation, we obtain the value
\begin{equation}
  X_{2,2} = X_t \simeq 0.3 \,.
\end{equation}

\begin{figure}[th]
  \begin{center}
    \includegraphics{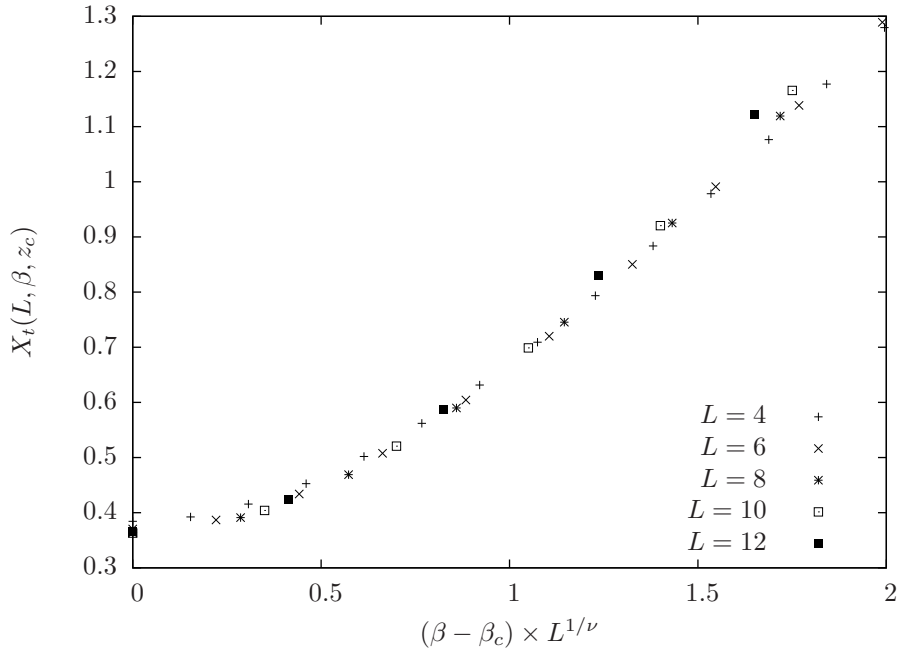}
    \caption{Data collapse for the effective thermal exponent $X_t(L,\beta,z_c)$,
      under a perturbation of the parameter $\beta$. The value used for this plot is $1/\nu=0.9$.}
    \label{fig:nu-trcc}
  \end{center}
\end{figure}

Finally, we evaluate the correlation-length exponent $\nu$ associated to
a perturbation of the parameter $\beta$ away from $\beta_c$. For the lowest
free-energy gap we expect the scaling form
\begin{equation}
  \log \frac{\Lambda_0}{\Lambda_1} \simeq \frac{2\pi}{L}
  \ F\left[ (\beta-\beta_c) \ L^{1/\nu}
  \right] \,.
\end{equation}
The best data collapse is obtained for the value (see Fig.~\ref{fig:nu-trcc}):
\begin{equation}
  \nu \simeq 1.1 \,.
\end{equation}


\subsection{Relation to Hilbert-space truncation}
\label{sec:Marston-etal}

We close this Section by comparing our approach to earlier studies~\cite{KM,MT}
of the IQHE problem based on a different truncation procedure.
Our method consists in writing the lattice path integral representation for
the mean conductance using the supersymmetry
trick, and then truncating the infinite sum over the paths, to keep only the
self-avoiding paths. This gives us the well-defined loop model of
Fig.~\ref{fig:cc-model}, where we will tune slightly the Boltzmann weights
to obtain an integrable point (see Section~\ref{sec:integrable}).

In contrast, in~\cite{KM,MT}, one starts from a two-dimensional single-particle
Hamiltonian including Gaussian hopping coefficients, and computes its
supersymmetric path integral.
The resulting action is then interpreted as the action of a one-dimensional
many-body supersymmetric Hamiltonian $H_{\rm MB}$, given in Eqs.~(3--5) of~\cite{MT}. This
Hamiltonian is expressed in terms of the coefficients $S^a$ of a superspin matrix.
In this model, the Hilbert space for each site is infinite-dimensional
(each site can be occupied by an arbitrary number of bosons).
The idea is to truncate this Hilbert space down to dimension $D$, and follow
the behaviour of the energy gap as $D$ increases. The model is not critical for finite
$D$, but it becomes critical in the limit $D \to \infty$.

Let us shown how to relate the terms of $H_{\rm MB}$ in the truncated space of dimension
$D=5$, to the generators which encode the loop model of Fig.~\ref{fig:cc-model}.
We first get rid of the $(-1)^j$ factor in $H_{\rm MB}$~\cite{MT}. This is
done through the change
\begin{equation*}
  c_{\up j} \to -c_{\up j} \qquad
  \hbox{for $j \equiv 2 \mod 4$ \quad or \quad $j \equiv 3 \mod 4$} \,,
\end{equation*}
without affecting the
(anti-)commutation relations for the $b_j,c_j$. We obtain the Hamiltonian:
\begin{equation}
  H_{\rm MB} = \sum_{j=1}^L \left[
    \sum_{a=1}^{16} g_a S^a_j S^a_{j+1}
    +\eta  (S^1_j + S^2_j + S^5_j + S^6_j)
  \right] \,,
\end{equation}
where the signs $g_a$ are given by
\begin{equation}
  g_a = \begin{cases}
    1 & \hbox{if $a = 1,2,10,12,14,16$} \\
    -1 & \hbox{if $a = 3,\dots,9,11,13,15$.}
  \end{cases}
\end{equation}
We decompose $H_{\rm MB}$ as a sum of generators
\begin{equation}
  H_{\rm MB} = \sum_{j=1}^{L} \Bigg\{
    -\dcup_j - \dcap_j + e_j + f_j
    + (1+\eta) \Big[ \idl_j + \idr_j + 2 \idlr_j \Big]
  \Bigg\} \,,
\end{equation}
where we have defined
\begin{equation}
  \begin{array}{rcl}
    \dcup_j &:=& S_j^3 S_{j+1}^3 + S_j^7 S_{j+1}^7
    + S_j^{15} S_{j+1}^{15} - S_j^{16} S_{j+1}^{16} \\
    \\
    \dcap_j &:=& S_j^4 S_{j+1}^4 + S_j^8 S_{j+1}^8
    + S_j^{13} S_{j+1}^{13} - S_j^{14} S_{j+1}^{14} \\
    \\
    \idl_j + \idlr_j &:=& \half \left( S^1_j + S^2_j + S^5_j + S^6_j \right) \\
    \\
    \idr_j + \idlr_j &:=& \half \left( S^1_{j+1} + S^2_{j+1} + S^5_{j+1} + S^6_{j+1} \right) \\
    \\
    e_j &:=& \left( S_j^1-\half \right) \left( S_{j+1}^1-\half \right)
    - \left( S_j^5+\half \right) \left( S_{j+1}^5+\half \right)
    + S_j^{10} S_{j+1}^{10} + S_j^{12} S_{j+1}^{12} \\
    \\
    f_j &:=& \left( S_j^2-\half \right) \left( S_{j+1}^2-\half \right)
    - \left( S_j^6+\half \right) \left( S_{j+1}^6+\half \right)
    - S_j^9 S_{j+1}^9 - S_j^{11} S_{j+1}^{11} \,.
  \end{array}
\end{equation}
In terms of the creation/annihilation operators, the above generators read:
\begin{equation}
  \begin{array}{rcl}
    \dcup_j &=& (b_{j\up}^\dag b_{j+1\up}^\dag + c_{j\up}^\dag c_{j+1\up}^\dag)
    (b_{j\dow}^\dag b_{j+1\dow}^\dag - c_{j\dow}^\dag c_{j+1\dow}^\dag) \\
    \dcap_j &=& (b_{j\up} b_{j+1\up} + c_{j\up} c_{j+1\up})
    (b_{j\dow} b_{j+1\dow} - c_{j\dow} c_{j+1\dow}) \\
    e_j &=& (b_{j\up}^\dag b_{j+1\up}^\dag + c_{j\up}^\dag c_{j+1\up}^\dag)
    (b_{j\up} b_{j+1\up} + c_{j\up} c_{j+1\up}) \\
    f_j &=& (b_{j\dow}^\dag b_{j+1\dow}^\dag - c_{j\dow}^\dag c_{j+1\dow}^\dag)
    (b_{j\dow} b_{j+1\dow} - c_{j\dow} c_{j+1\dow}) \\
    \idl_j + \idlr_j &=& \half (b_{j\up}^\dag b_{j\up} + c_{j\up}^\dag c_{j\up}
    + b_{j\dow}^\dag b_{j\dow} + c_{j\dow}^\dag c_{j\dow}) \\
    \idr_j + \idlr_j &=& \half (b_{j+1\up}^\dag b_{j+1\up} + c_{j+1\up}^\dag c_{j+1\up}
    + b_{j+1\dow}^\dag b_{j+1\dow} + c_{j+1\dow}^\dag c_{j+1\dow}) \,.
  \end{array}
\end{equation}
In the $D=5$ truncated space, each site is either empty or occupied by two particles of
opposite spins $(\up,\dow)$. If each spin is interpreted as a loop color, the above generators
(when restricted to the $D=5$ space) obey a dilute two-color Temperley-Lieb algebra with loop
weight $n=0$. Hence, they represent the vertices $u_1, u_2, w_1, x$ of the loop model defined in
Section~\ref{sec:truncation}. In particular, the $e_j$ and $f_j$ form two decoupled
Temperley-Lieb algebras.

Note that, in this context, the generator for the $w_2$ vertex, $E_j=e_j f_j$,
cannot be realised by a linear combination of the $S^a_j S^a_{j+1}$, but it may be
a linear combination of the $(S^a_j S^a_{j+1})^2$. So introducing $E_j$ terms in the Hamiltonian
leads to higher-order terms in $H_{\rm MB}$, and most probably it breaks the invariance
with respect to the supersymmetric charges $Q_{1,2}$.
However, we have shown that the supersymmetric model $H_{\rm MB}$, when restricted to the
$D=5$ space, corresponds to a particular manifold in the phase diagram of the two-color
loop model.

\section{Construction of an integrable critical loop model}
\label{sec:integrable}

\begin{figure}[th]
  \begin{center}
    \includegraphics{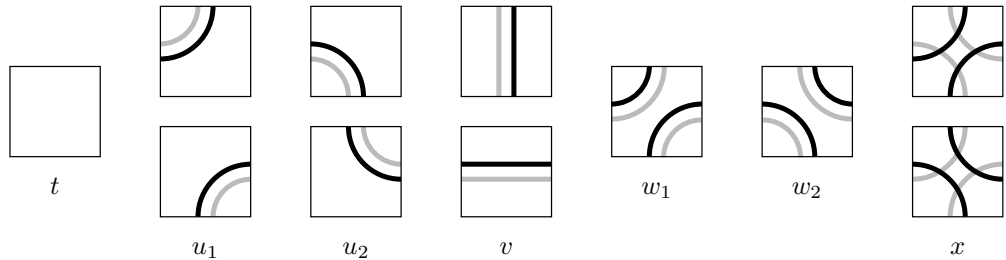}
    \caption{Vertices of the augmented dilute two-color loop model.}
    \label{fig:model}
  \end{center}
\end{figure}

In the preceding section, we truncated the the Chalker-Coddington
network model to yield a two-color loop model that is simpler to
analyze. To make further progress, we modify this model further. We
augment it by allowing the ``straight-line'' vertices with weight $v$
illustrated in Fig.~\ref{fig:model}. We also generalize it by
allowing the weight per loop $n$ to not only be zero, but to vary in
the range $n \in [-2,2]$. By utilizing the results of \cite{GW},
we will show in this section that
this modified model for all values of $n$ in this range
has an integrable line, and includes several critical points.
The remainder of the paper will be devoted to
the study of the critical behaviour.

When the straight-line vertices are allowed, the loop model can no
longer be related directly to electron trajectories in a potential.
In the original CC model, the `checkerboard' structure of the lattice
(or, equivalently, the alternation of arrows on the edges of $\cal L$)
is essential to the interpretation of the paths as electron
trajectories along the contour lines of the random potential. However,
several arguments indicate that the truncated but unmodified loop model
of Fig.~\ref{fig:cc-model} is in the same universality class as that
of the modified model. In other words, one can obtain the unmodified
model by perturbing the critical line with irrelevant operators.

One argument for the equivalence of the two stems from the relation of
this two-color loop model to that studied in \cite{Paul-Jesper}. There
the completely packed version was studied; in the notation used here
this corresponds to setting the Boltzmann weights $t= u_1= u_2= v= 0$. It
was shown that at least for weight per loop $n\ge \sqrt{2}$, the model
has a critical point when $x/w_2$ is tuned appropriately. Moreover, at
this critical point, numerical evidence strongly suggests that
dilution (i.e.\ non-zero $t$, $u_1$ and $u_2$) is irrelevant. We will
provide additional evidence by finding that for certain discrete
values of $n\ge \sqrt{2}$, the critical point of the completely packed
model and that of the modified model studied here are described by the
same conformal field theory. Neither of these arguments
applies when $n=0$, but all the critical exponents we have computed ($X_t, X_{2,2}, \nu$)
for both the truncated CC model and the integrable model at $n=0$ in regime 4
(see Table~\ref{table:regimes}) agree, up to our numerical precision.
This strongly indicates that the integrable model at $n=0$ in regime 4
is in the universality class of the truncated CC model.

In this section, we give the Boltzmann weights of the integrable
critical line in the loop model of Fig.~\ref{fig:model}. These weights
are expressed in terms of the generators of the dilute
Birman-Wenzl-Murakami (dBWM) algebra, so that the solution of the
Yang-Baxter equation found in \cite{GW} can be used. In Appendix~A,
we review the BWM algebra and its graphical
presentation. The braid group can be represented in terms of the BWM
generators, and can then be used to find invariants of knots and links
generalizing the Jones polynomial \cite{BWM}.

An alternate way of obtaining the Boltzmann weights of the integrable critical
line is to search for holomorphic observables on the lattice. These
are operators whose expectation values satisfy the lattice analog of
the Cauchy-Riemann equations. This method is described in Appendix B,
and yields the same weights as those found in \cite{GW} using the dBWM
algebra.

\subsection{Critical completely packed loop models}
\label{sec:CPL}

\begin{figure}[ht]
  \begin{center}
    \includegraphics[scale=0.8]{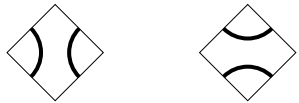}
  \end{center}
  \caption{Action of $\idop$ (left) and $e_j$ (right) on a pair of strands
    at positions $j$ and $j+1$. The transfer matrix direction is
    upwards.}
  \label{fig:TL}
\end{figure}

We first review the critical completely packed loop model,
arising for example in the Fortuin-Kasteleyn expansion of the Potts
model \cite{FK}. Each vertex of this model has the two possible
configurations displayed in Fig.~\ref{fig:TL}.
The partition function is conveniently written
in terms of the generators of the Temperley-Lieb (TL) algebra \cite{TL}.
This algebra for a system of width $L$ has
$L$ generators $e_j$ acting at positions $j=1,2,\ldots,L$ as well as the
identity $\idop$, which obey the relations
\begin{equation} \label{TLalg}
  e_j^2 = n \ e_j \,,
  \qquad e_j e_{j\pm 1}e_j = e_j \,,
  \qquad e_i e_j = e_j e_i \quad\hbox{for $|i-j|>1$.}
\end{equation}
The first of the relations encodes the fact that the weight for a
closed loop is $n$, while the second encodes the fact that
the weight does not depend on the length or the shape of the loop.

The Boltzmann weights of the integrable critical loop model are then
\begin{equation}
  \Rc_j(u)= \sin(2\theta-u) \, \idop - \sin u \, e_j \,,
\end{equation}
where $n=-2\cos 2\theta$ and $|n|\le 2$.
The transfer matrix for an even number of sites $L$ is then
\begin{equation}
  T = \Rc_1 \Rc_3 \dots \Rc_{L-1} \Rc_2 \Rc_4 \dots \Rc_L \, .
\end{equation}
It is straightforward to use the TL algebra to verify that
these Boltzmann weights satisfy the Yang-Baxter equation
\begin{equation}
  \Rc_j(u) \Rc_{j+1}(u+v) \Rc_j(v) = \Rc_{j+1}(v) \Rc_j(u+v) \Rc_{j+1}(u)
\end{equation}
and the inversion relation
\begin{equation}
  \Rc_j(u) \Rc_j(-u) = \sin(2\theta-u) \sin(2\theta+u) \ \idop \,.
\end{equation}
Braid group generators $b_j$ and $b_j^{-1}$ are found by taking $u \to \pm i\infty$:
$$
\Rc_j (i\infty)\propto b_j = e^{-i\theta}\, \idop + e^{i\theta}\, e_j,
\qquad
\Rc_j (-i\infty)\propto b^{-1}_j = e^{i\theta}\, \idop + e^{-i\theta}\, e_j \ .
$$
These satisfy the braid-group relations \eqref{eq:braid1} and
\eqref{eq:braid2} as a consequence of the Yang-Baxter equation and
the inversion relation respectively.

The critical completely packed loop model on the square lattice is in
the same universality class as what is usually known as the $\On$
model in its dense phase. Well-established results on the dense $\On$
model~\cite{nienhuis-On} give the central charge of the CFT
describing the scaling limit to be
\begin{equation}
  c_{\On} = 1 - \frac{3(\pi -2\theta)^2}{\pi\theta} \ .
\label{eq:cOn}
\end{equation}

The Boltzmann weights of the completely packed doubled loop model
studied in \cite{Paul-Jesper,Martins-Nienhuis} can be written in terms
of the generators $e_i$ and $f_i$ of two independent TL
algebras. This model is displayed in Fig.~\ref{fig:model} with $t=u_1=
u_2=v=0$. In this picture, the $e_j$ acts on black loops while the
$f_j$ act on grey loops, while the transfer matrix goes to the
northeast. Thus the vertex with weight $w_1$ corresponds to the
generator $\idop$, the vertex with weight $w_2$ corresponds to $e_j f_j$,
while those with weight $x$ are $e_j$ and $f_j$. Since the $e_j$'s and
the $f_j$'s commute, we have immediately that the $B_j, B^{-1}_j$
defined by
\begin{equation} \label{eq:def-Bj}
  \begin{array}{rcl}
    B^{\phantom{-1}}_j &:=& \left(e^{-i\theta} \, \idop + e^{i\theta} \, e_j \right)
    \left(e^{-i\theta} \, \idop + e^{i\theta} \, f_j \right)\\
    B^{-1}_j &:=& \left(e^{i\theta} \, \idop + e^{-i\theta} \, e_j \right)
    \left(e^{i\theta} \, \idop + e^{-i\theta} \, f_j \right)
  \end{array}
\end{equation}
also generate a braid group.
Similarly, TL generators with loop weight $N=n^2$
may be constructed as
\begin{equation}
  E_j := e_j \ f_j \,.
\end{equation}

Using the relations (\ref{TLalg}) for the $e_j$'s and $f_j$'s, it is
straightforward to show that the $B_j, B^{-1}_j, E_j$ generate the BWM
algebra described in Appendix~A with parameters
$N=n^2=(-2\cos 2\theta)^2$, $\omega=e^{i6\theta}$
~\cite{Paul-Jesper}. The doubled lines here correspond to the single
lines displayed in Appendix~A, as is apparent by comparing
Figs.~\ref{fig:model} and \ref{fig:dBWM-gen}. Writing the Boltzmann weights
in terms of this algebra is useful because solutions of the
Yang-Baxter equation involving the BWM generators have long been known
\cite{Wadati}. From this solution, a critical point for the coupled
completely packed loop models for $n\ge \sqrt{2}$ was found
\cite{Paul-Jesper,Martins-Nienhuis}. 
With the parameterisation
$$
n= 2\cos \frac{\pi}{r+2} \,,
$$
in the isotropic case $w_1=w_2$,
the critical point is at  $x/w_1 = \lambda_c$, where
\begin{equation}
  \lambda_c =
  -\sqrt{2} \sin \left[ \frac{\pi(r-2)}{4(r+2)} \right] \,.
\label{eq:lambdac}
\end{equation}
At integer values $r=2,3,4 \dots$, this critical point was
identified with a particular
conformal field theory, the WZW coset model $\SU(2)_r \times \SU(2)_r/\SU(2)_{2r}$.
This conformal field theory has central charge
\begin{equation}
  c_r = \frac{3r^2}{(r+1)(r+2)}\ .
  \label{eq:cr}
\end{equation}
For $1/\lambda_c < x/w_1 < \lambda_c$, the doubled loop model has a critical
phase corresponding to two decoupled completely packed loop
models. The central charge is thus twice (\ref{eq:cOn}).

\subsection{The integrable critical line}
\label{sec:integrable-line}

We now can use the results of Grimm and Warnaar \cite{GW} to find an
integrable model involving all the vertices in Fig.\
\ref{fig:model}. We are interested mainly in the critical points, which
can be written in terms of the dilute BWM algebra. The dilute BWM algebra extends
the BWM algebra described in Appendix~A to include edges of the lattice uncovered by
strands. In the two-color loop model, these amount to allowing vertices to
be empty of both colors. The dilute generators act identically on the
two colors, and so include the remaining vertices in Fig.\
\ref{fig:model}. In an obvious notation, we then can write the
$\Rc$-matrix as
\begin{eqnarray}
  \Rc_j(\varphi) &=& t(\varphi) \vac_j
  + u_1(\varphi) \big[ \idl_j + \idr_j \big]
  + u_2(\varphi) \big[ \dcup_j + \dcap_j \big] \nonumber \\
  && + v(\varphi) \big[ \vup_j + \vdown_j \big]
  + w_1(\varphi) I_j + w_2(\varphi) E_j + x(\varphi) X_j \,.
  \label{eq:R-mat}
\end{eqnarray}
In terms of the TL generators introduced in the previous
section, $E_j=e_j f_j$ and $X_j \equiv e_j + f_j$, while $I_j$ takes value $0$ on the
dilute configurations and $1$ otherwise.

Since the non-dilute vertices satisfy the BWM algebra, it is simple to
show that the operators
$$
B_j, E_j, I_j, (\ \ )_j, \idl_j, \idr_j, \dcup_j, \dcap_j,
\vup_j, \vdown_j
$$
constructed from the two-color loop model satisfy a dilute BWM algebra.
Namely, with doubled lines here corresponding to single lines in
Appendix~A, and the $B_j$ defined in \eqref{eq:def-Bj}, these operators
generate the dilute BWM algebra with parameters $(N=(q+q^{-1})^2,
\omega=q^3)$, where $q=e^{2i\theta}$.

In~\cite{GW}, an integrable model based on the dBWM algebra was derived.
With $n=-2 \cos 2\theta$ as before, its Boltzmann weights are given by
\begin{equation} \label{eq:weights}
  \begin{array}{rcl}
    t(\varphi)   &=& -\cos(2\varphi-3\theta) - \cos 5\theta + \cos 3\theta + \cos \theta \\
    u_1(\varphi) &=& -2 \sin 2\theta \sin(\varphi-3\theta) \\
    u_2(\varphi) &=& \phantom{-} 2 \sin 2\theta \sin \varphi \\
    v(\varphi)   &=& -2 \sin \varphi \sin(\varphi-3\theta) \\
    w_1(\varphi) &=& \phantom{-} 2 \sin(\varphi-2\theta) \sin(\varphi-3\theta) \\
    w_2(\varphi) &=& \phantom{-} 2 \sin \varphi \sin(\varphi-\theta) \\
    x(\varphi)   &=& \phantom{-} 2 \sin \varphi \sin(\varphi-3\theta) \,.
  \end{array}
\end{equation}
We denote by $\varphi_0$ the isotropic value which is closest to zero: 
\begin{equation} \label{eq:phi0}
  \varphi_0 =\begin{cases}
    \frac{3\theta}{2}
    & \hbox{if $0< \theta < \frac{\pi}{3}$} \\
    \frac{3\theta}{2}-\pi
    & \hbox{if $\frac{\pi}{3} < \theta < \pi$.}
  \end{cases}
\end{equation}
The universal properties are independent of the anisotropy parameter
$\varphi$ (as long as $\varphi$ lies between $0$ and $\varphi_0$), but depend very strongly on
$\theta$, as we shall see.  At the isotropic point
$\varphi=3\theta/2$, the weights can be rescaled to
\begin{equation}
  \begin{array}{l}
    t = 2\cos 3\theta + 2 \cos 2\theta + 1 \\
    u_1=u_2 = 4 \cos \frac{\theta}{2} \cos \theta \\
    v = 2\cos \theta + 1 \\
    w_1=w_2 = 1 \\
    x = -(2\cos \theta + 1) \,.
  \end{array}
\end{equation}

The integrable model defined by \eqref{eq:R-mat}--\eqref{eq:weights} obeys the
following properties:
\begin{itemize}

\item The isotropic weights are invariant under the transformations
  $\theta\to 2\pi+\theta$, and $\theta\to -\theta$, so the range of
  inequivalent couplings is $\theta\in [0,\pi]$. Each value of $n\in [-2,2]$
  appears twice in this interval.

\item Since there are no loop ends, the
  number of loops mod 2 is the same as the number of $x$ vertices mod
  $2$. This allows us to change the sign of $n$ by absorbing the sign
  in the weight $x$: $(n,x) \to (-n,-x)$. Thus there are four distinct
  critical points for each value of $n\in (0,2)$, while there are two
  for $n=0$ and $n=2$.

\item The weights satisfy the inversion relation
  \begin{equation}
    \label{eq:inversion}
    \check{R}(\varphi)\check{R}(-\varphi)
    = 4 \sin(2\theta-\varphi)\sin(2\theta+\varphi)
    \sin(3\theta-\varphi)\sin(3\theta+\varphi) \ \idop \,.
  \end{equation}

\item Rotating by 90$^{\rm o}$ is equivalent to sending
  $\varphi \to 3\theta-\varphi$.

\item The weights are trivial when $u=0$:
  $\check{R}(0)= 2 \sin 2\theta \sin 3\theta \ \idop \,.$

\item The eigenvalues of the transfer matrix are preserved under
  $(u_1,u_2) \to (-u_1,-u_2)$ and $v \to -v$.

\end{itemize}

\subsection{The quantum Hamiltonian}
\label{sec:Hamiltonian}

To gain intuition into this doubled loop model, it is useful to find
the equivalent 1d quantum Hamiltonian by taking the very anisotropic
limit $\varphi \to 0$. The Hamiltonian is found from the transfer
matrix $T_L(\varphi)$ for $L$ sites by
$$
  H := 2\sin 2\theta \sin 3\theta \ \left.
    \frac{{\rm d}\log T_L(\varphi)}{{\rm d}\varphi}
  \right|_{\varphi=0}
  + 2L \sin 5\theta \ \idop \ ,
$$
yielding
\begin{eqnarray}
H &=& \sum_{j=1}^L \Big\{
  4 \cos 4\theta \sin \theta \ \vac_j
  + 2 \cos 2\theta \sin 3\theta \ \left[ \idl_j + \idr_j \right]
  + 2 \sin 2\theta \ \left[ \dcup_j + \dcap_j \right] \nonumber \\
  && \qquad +2 \sin 3\theta \ \left[ \vup_j + \vdown_j \right]
  -2 \sin \theta \ E_j -2 \sin 3\theta \ X_j
  \Big\} \,.
\label{eq:ham}
\end{eqnarray}
To find the Fermi velocity $v_f$, we assume that in the scaling limit
this Hamiltonian is that of a conformal field theory. In the next
Section, we will present much evidence in support of this assumption.
In a conformal field theory, the ground-state energy
(the lowest eigenvalue of $H$) is \cite{E-CFT}
\begin{equation} \label{eq:E-CFT}
  E^0_L \simeq L e_\infty
  -\frac{\pi c}{6L} \ v_f \,.
\end{equation}
where $c$ is the central charge. Let $\Lambda^0_L(\varphi)$ be the
dominant eigenvalue of the transfer matrix. The analysis of Appendix~B
indicates that the free energy of the loop model on a rhombic
lattice with angle $\alpha$ is given by $\left[-\log
\Lambda^0_L(\varphi) \right]$, where $\alpha = \pi\varphi/(2\varphi_0)$ and $\varphi_0$ is the isotropic value, as
defined in \eqref{eq:phi0}.
In a conformal field theory, one expects \cite{E-CFT}
\begin{equation} \label{eq:Lambda-CFT}
  - \log \Lambda^0_L(\varphi)\simeq  L f_\infty(\alpha)
  -\frac{\pi c}{6L} \sin \alpha \,.
\end{equation}
Differentiating \eqref{eq:Lambda-CFT} around $\varphi=0$ and
comparing with \eqref{eq:E-CFT} yields
\begin{equation} \label{eq:vf}
  v_f = \left|
    \frac{2\pi \sin 2\theta \sin 3\theta}{2\varphi_0}
  \right| \,,
  \qquad \varphi_0 =\begin{cases}
    \frac{3\theta}{2}
    & \hbox{if $0< \theta < \frac{\pi}{3}$} \\
    \frac{3\theta}{2}-\pi
    & \hbox{if $\frac{\pi}{3} < \theta < \pi$.}
  \end{cases}
\end{equation}

\section{Identifying the critical theories}
\label{sec:critical-th}

In this Section, we present what we believe is convincing evidence that
the doubled loop model with Boltzmann weights~\eqref{eq:weights} is
critical. We find the presumably exact central charge of the conformal
field theories describing the scaling limit, and also give some of the
dimensions of fields. We do this by a combination of calculations
exploiting the integrability, comparison to a similar integrable
model, and exact diagonalization of the transfer matrix and the Hamiltonian
for widths up to $L=14$ sites.

\subsection{The four regimes}
\label{sec:regimes}

This critical line is parametrized by the value of $\theta\in[0,\pi]$,
related the weight per loop by $n=-2\cos 2\theta$. Since the Fermi
velocity vanishes at $\theta=\frac{\pi}{3}, \frac{\pi}{2}$, and has a discontinuity at
$\theta=\frac{2\pi}{3}$, it is natural to expect that the physics is
discontinuous if $\theta$ is varied across these values. We thus
divide the critical line into four regimes, as described in
Table~\ref{table:regimes}.

\begin{figure}[th]
  \begin{center}
    \begin{tabular}{cc}
      \includegraphics[width=.4\textwidth]{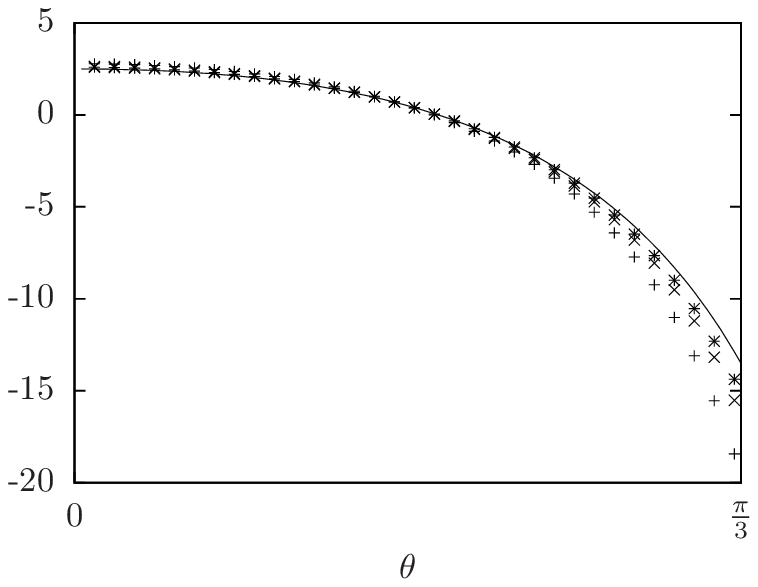} &
      \includegraphics[width=.4\textwidth]{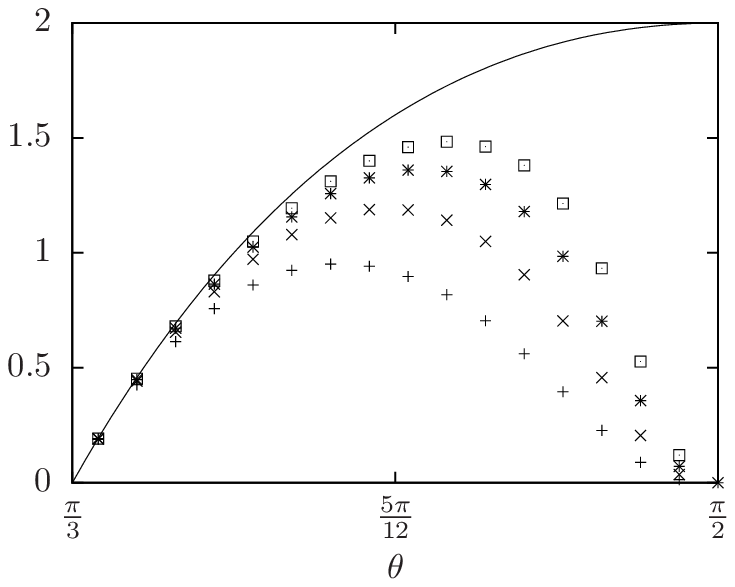} \\
      \includegraphics[width=.4\textwidth]{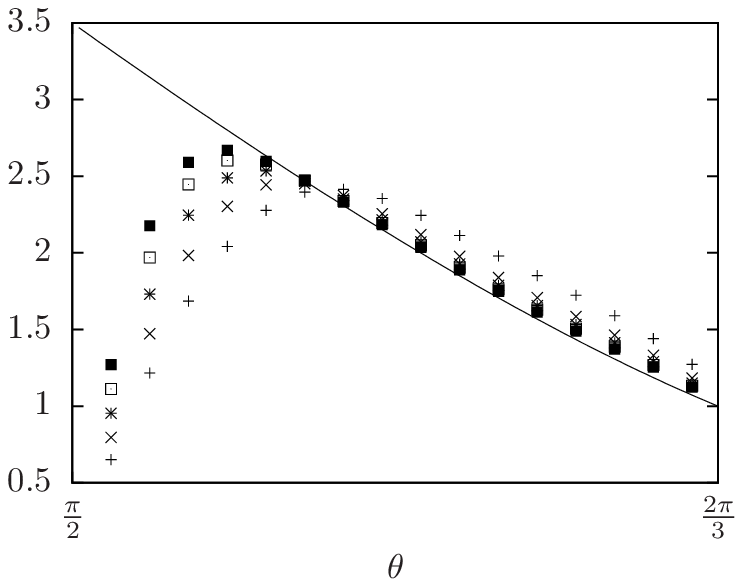} &
      \includegraphics[width=.4\textwidth]{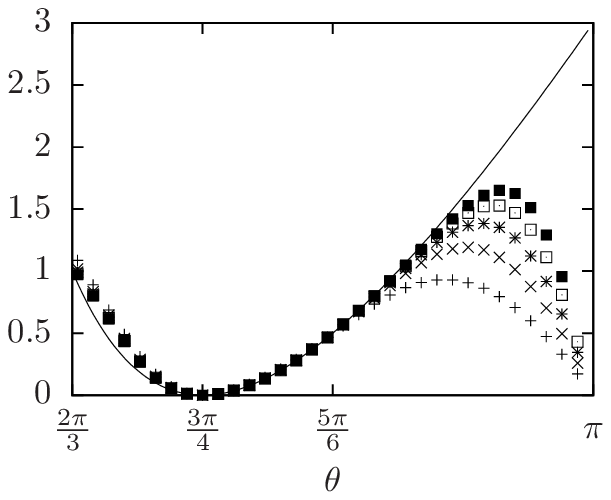}
    \end{tabular}
    \caption{Numerical estimates for the central charge in the four
      critical regimes. Different symbols represent data points for
      consecutive system sizes:
      $L=4,6 \ (+), L=6,8 \ (\times), L=8,10 \ (*), L=10,12 \ ({\scriptstyle \square}), L=12,14 \ ({\scriptstyle \blacksquare})$.
      Full lines represent the predicted
      exact values from Table~\ref{table:regimes}.}
    \label{fig:cc}
  \end{center}
\end{figure}

All known integrable models with Boltzmann weights parameterized by
trigonometric functions of the anisotropy parameter $\varphi$ are
critical, and this is no exception. One argument for this is the
existence of the lattice holomorphic operator described in Appendix B.
Another is the inversion-relation calculation done below, which shows
that with standard assumptions about holomorphicity in $\varphi$, the free
energy is singular as this critical point. A numerical check is to use
exact diagonalization to find the largest eigenvalue of $T$ and/or the
ground-state energy of $H$, and then fit the results to \eqref{eq:E-CFT} or
\eqref{eq:Lambda-CFT}. To extract the central charge $c$, we use two
different-length systems to get rid of the extensive piece
$L e_\infty$. Doing this, we find the results given in
Fig.~\ref{fig:cc}. We see a very nice convergence to the
critical behavior as expected.

\begin{table}
  \begin{center}
    \begin{tabular}{|c|c|c|c|}
      \hline
      {\bf regime} & {\bf $\theta$-range} & {\bf parameterisation}
      & {\bf central charge} \\ \hline
      & & & \\
      1 & $0<\theta<\frac{\pi}{3}$ & $n=-2\cos \frac{\pi}{r+2}$ &
      $c=2 \left[ 1 - \frac{6}{(r+1)(r+2)} \right]+ \half$ \\
      & & & \\ \hline
      & & & \\
      2 & $\frac{\pi}{3} < \theta < \halfpi$ & $n= \phantom{-} 2\cos \frac{\pi}{r+2}$ &
      $c=2 \left[1 - \frac{6}{(r+1)(r+2)} \right]$ \\
      & & & \\ \hline
      & & & \\
      3 & $\halfpi < \theta < \frac{2\pi}{3}$ & $n= \phantom{-} 2\cos \frac{\pi}{r+2}$ &
      $c= \frac{3 r^2}{(r+1)(r+2)} + \half$ \\
      & & & \\ \hline
      & & & \\
      4 & $\frac{2\pi}{3} < \theta < \pi$ & $n= -2\cos \frac{\pi}{r+2}$ &
      $c= \frac{3 r^2}{(r+1)(r+2)}$ \\
      & & & \\ \hline
    \end{tabular}
    \caption{The four regimes of the integrable loop model.}
    \label{table:regimes}
  \end{center}
\end{table}

We combine these results with other arguments to conjecture
exact formulae for the central charge for all $\theta$.  We can also
identify precisely which conformal field theories describe some
critical lines.  There are two types of conformal field theories known
to describe doubled loop models, and both occur along
this critical line. Unfortunately, the value of $n=0$ at
$\theta= \frac{3\pi}{4}$ of interest for the truncated CC model lies in one of
the regions where we do not understand the conformal field theory. As
is apparent from Fig.~\ref{fig:cc}, we do know that $c=0$ as required
there.

At several special values of $\theta$, the model simplifies. Namely,
when $n=\pm 1$, all loop configurations receive the same weight (if
$n=-1$, we transform $(n,x) \to (-n,-x)$ as explained in Sec.~\ref{sec:integrable-line}).
Thus when computing the
partition function, we can sum up the four completely packed vertices
to give a single one with weight $w_1+w_2+ 2nx$. 

For $\theta=\frac{2\pi}{3}$ at the isotropic point, $x=v=0$, so this
reduces to a six-vertex model with no staggering. Here the usual
parameter~\cite{Baxbook} has value
$$
\Delta= \frac{a^2+b^2-c^2}{2ab} = -1 \,,
$$
so this is in the same universality class as the antiferromagnetic
Heisenberg model. Thus the central charge is $c=1$ and the 
first thermal exponent is $X_t= \half$,
in agreement with the numerical results in Figs.~\ref{fig:cc}--\ref{fig:xt}.

At $\theta=\frac{\pi}{6}$ and
$\theta=\frac{5\pi}{6}$, we obtain a staggered version of the eight-vertex
model. Ordinarily the staggered eight-vertex model is not solvable,
but as we detail in Appendix~C, this one is not only solvable, but can
be mapped onto a free-fermion theory. There we show that there are two
Majorana fermions present, but only one of the two is critical. Thus
the central charge is $c=\half$ here, again consistent with the numerics.

\subsection{Computation of an exact scaling dimension}
\label{sec:exact-dim}

Since the model is integrable, it is possible to derive some
quantities exactly. Here we extract the dimension of an operator in
the critical theory as a function of $\theta$. This is possible
because at certain discrete values of $\theta$, there exists a
deformation away from the critical point preserving the
integrability \cite{GW}.
The inversion-relation method \cite{Baxbook} yields the
free energy along this deformation, and by analyzing its expansion
around the critical point, we extract the value of the exponent
$\nu_{\rm int}$. This then yields the dimension of the operator which
when added to the action causes the deformation.

It is convenient to parameterize the loop weight $n$ within each
of the four regimes by a parameter~$r$,
\begin{equation}
  n= 2 \epsilon \cos \frac{\pi}{r+2} \,,
\end{equation}
where $\epsilon=-1$ in regimes 1 and 4, and $\epsilon=1$ in regimes 2 and 3
(see Table~\ref{table:regimes}).

The integrable deformations resulting in unitary field theories occur
at integer values of $r$ in all four regimes. Here the dilute BWM algebra
admits a ``height'' or ``Restricted Solid-On-Solid'' (RSOS)
realization~\cite{GW}. Instead of treating the loops as the degrees of
freedom, on the dual lattice one places height variables, which are
integers restricted to a certain interval. The loops then play the
role of domain walls separating regions of different heights.

The inversion-relation method is a way of computing the free energy
exactly after making assumptions about its holomorphicity
properties as a function of $\varphi$. The free energy satisfies
constraints following from the inversion relation
\eqref{eq:inversion2} below, and the fact that sending $\varphi\to
3\theta-\varphi$ rotates the lattice by $90^{\rm o}$. The holomorphicity
assumptions then give a unique solution to these
constraints. Parameterizing the deformation in our case by $p$,
the inversion relation becomes \cite{GW}
\begin{equation}
  \label{eq:inversion2}
  \check{R}(\varphi,p)\check{R}(-\varphi,p)
  = (4p)^{-1}\, \theta_1(2\theta-\varphi,p)\,\theta_1(2\theta+\varphi,p)\,
  \theta_1(3\theta-\varphi,p)\,\theta_1(3\theta+\varphi,p) \ \idop \,,
\end{equation}
where $\theta_1(u,p)$ is the standard elliptic theta function. This
indeed reduces to (\ref{eq:inversion}) in the critical limit $p\to
0$. From this, it is simple to show that the inverse of the transfer
matrix in the diagonal direction is given by forming a transfer
matrix out of products of $\Rc_j(-\varphi,p)$.

Conveniently, both (\ref{eq:inversion2}) and the behavior under
rotational symmetry are identical to that of the model studied in
\cite{WPSN}, so we utilize these results. The singular part of the
free energy approaching the critical point depends on $p$ as $f_{\rm
sing} \sim p^{2-\nu_{\rm int}}$, so the operator perturbing the
critical theory in the integrable direction has scaling dimension
$X_{\rm int}=2-2/\nu_{\rm int}$. Then we find
\begin{equation} \label{eq:xint}
  X_{\rm int} = \begin{cases}
    \frac{r-1}{r+2}+1 & \qquad \hbox{in regime\ 1,}\\
    \frac{r-1}{r+2}   & \qquad \hbox{in regime\ 2,}\\
    \frac{3}{r+2}+1   & \qquad \hbox{in regime\ 3,}\\
    \frac{3}{r+2}     & \qquad \hbox{in regime\ 4.}
  \end{cases}
\end{equation}
We have written these results in terms of $r$ instead of $\theta$ to
emphasize that the derivation only applies to $r$ integer, since this
is where (\ref{eq:inversion2}) can be derived. However, we
expect that the results can be continued to all $r$ within a given
regime, since the equations themselves depend on $r$ as a continuous parameter.

A good check of the validity of~\eqref{eq:xint} for generic $\theta$ is that
it corresponds exactly to $X_{\rm int} = 2s$, where $s$ is the spin of the
discretely holomorphic parafermion $\psi_s(z)$ described in Appendix~B:
see~\eqref{eq:s}.

\subsection{Description by conformal field theory}
\label{sec:CFT}

Here we give formulae for the central charges in all four regimes that
are presumably exact. All are related to those occurring in
completely packed models. However, the doubled loop models are not
identical to completely packed models: we have checked that
the doubled loops have non-trivial fractal
dimension in regime 4 (see Section~\ref{sec:other-exp}).

One type of critical behavior possible for a doubled loop model is
simply to have the two colors decouple in the scaling limit. This
occurs in the isotropic completely packed version when $x=-w_1=-w_2$,
and is argued to persist in a region around this point
\cite{Paul-Jesper}. The central charge is simply twice $c_{O(n)}$
given in (\ref{eq:cOn}).  Examining our numerical results for $c$, we
see that in regime 2, the central charge indeed is converging nicely
to $2c_{O(n)}$ with the appropriate dependence on $n$. When $r$ is an
integer, this is twice the central charge of the conformal minimal
models, and so the corresponding height model should scale to two
decoupled minimal models. An additional check on this comes from the
fact that the dimension of the integrable perturbing operator in
(\ref{eq:xint}) is twice that of an operator in a minimal
model. Namely, we have $X_{\rm int}=2X_{1,2}$, where $X_{1,2}=
\frac{r-1}{2(r+2)}$ is the scaling dimension of the $\Phi_{1,2}$ operator in
the minimal model with central charge $c=1- \frac{6}{(r+1)(r+2)}$.
It is thus natural to conjecture that in regime 2, the scaling limit
of our integrable loop model is indeed that of two decoupled
completely packed loop models. These conformal field theories have
been extensively studied~\cite{minimal}.

In regime 1, the numerics for the central charge are apparently
converging to $2c_{O(n)}+\half$. Thus here the loops apparently decouple as
well, but additional critical Ising degrees of freedom appear.  This
is consistent with the mapping to the Ising model at $\theta=\frac{\pi}{6}$ described in
Appendix~C. The scaling dimension $X_{\rm int}$ here is $2X_{1,2}+1$,
leading to the natural interpretation that the operator is a product
of the $\Phi_{1,2}$ operators in the two minimal models with the
energy operator in the Ising model, the latter having dimension $1$.
These extra Ising degrees of freedom, which also occur in
a certain regime of the square-lattice $\On$ model~\cite{square-On},
appear through the following mechanism.  The vertices of the loop
model obey a $\mathbb{Z}_2$ symmetry, in the sense that any vertex is
surrounded by an even number of empty edges. Thus empty edges form
polygons where each node has even degree, and so they respect the
geometry of Ising domain walls for Ising variables $\sigma$ lying on
the dual lattice. Depending on the values of the Boltzmann weights,
these Ising variables may become critical in the continuum limit. This
is evidently what happens in regime 1.

The critical behavior in regimes 3 and 4 is not that of two decoupled
models. As mentioned above in Sec.~\ref{sec:CPL}, in the
completely packed version of the doubled loop model, there occurs a
coupled critical point corresponding to the $\SU_r(2) \times \SU_r(2) / \SU_{2r}(2)$ WZW
coset model, with central charge $c_r$ \eqref{eq:cr}.
The numerical analysis in Fig.\ \ref{fig:cc} nicely fits to
$c_r$ in regime 4, and agrees with the Ising value $c=\half$ at
$\theta=\frac{5\pi}{6}$, derived in Appendix~C. Moreover, when $r$ is an
integer, the exponent $X_{\rm int}$ in (\ref{eq:xint}) belongs to the
above coset theory. Thus it is natural to conjecture that the central
charge throughout regime 4 is $c_r$.
As we see from our numerics at $n=0$ (see Sec.~\ref{sec:other-exp}), 
the fractal dimension of a single loop is $d_f<2$, so regime
4 represents a ``dilute branch'' of the coset theory.

Likewise, in regime 3 the data seem to be converging to $c_r+\half$,
agreeing with the six-vertex value $c=1$ at $\theta=\frac{2\pi}{3}$. The
exponent $X_{\rm int}$ is 1 greater than the value in regime 4, so it
is natural to interpret that the operator is multiplied by the Ising
energy operator of dimension of 1. Thus like in regime 1, the critical
theory presumably includes an extra Ising piece.

Outside $r$ integer, the conformal field theory in regimes 3 and 4 is
not understood. Moreover, we will see in the subsequent section that
even though the formula for the central charge is applicable for all
$r$, it is not even clear whether dimensions of exponents can be
continued to values of $|n|<1$.

\section{Critical behaviour in regime 4}
\label{sec:regime4}

The main motivation of this paper is to explore a doubled loop model
arising in the truncation of the Chalker-Coddington network model. For
a connection to disordered systems, the weight per loop $n$ and the
central charge $c$ must be zero. We have two $n=0$ points, but for
$\theta=\frac{\pi}{4}$ inside regime 1, the corresponding critical field
theory seems to have nothing to do with the CC model. Not only do the
different colors of loop decouple, but the extra Ising degree of
freedom makes $c\ne 0$. We thus in this section focus on the behavior
in regime 4, which contains the other $n=0$ point at $\theta=\frac{3\pi}{4}$.

As noted above, we do not have a conformal field theory description
valid in regime 4 outside of integer $r$. It therefore seems a good
idea to exploit the fact that the associated height
description at these points is described by the coset conformal field
theory $\SU(2)_r \times \SU(2)_r / \SU(2)_{2r}$.

\subsection{Integrable perturbations}
\label{sec:integrable-pert}

This coset theory is known to have two integrable perturbations. One
of them, found by using level-rank duality on the results of~\cite{Vaysburd},
is by the operator with dimension $X_{\rm int}=\frac{3}{r+2}$ discussed
above. This perturbation describes the scaling limit of the height
model with elliptic Boltzmann weights~\cite{GW}. In terms of the loop
model, we have found that the discrete parafermion~$\psi_s(z)$ of Appendix~B
is the chiral part of the corresponding operator. This parafermion consists
of the insertion of a one-leg defect for each loop flavour.

The other integrable perturbation also has a very natural
meaning in terms of loops. This perturbing operator corresponds
to the (1,1;adjoint) operator of dimension $X_{(1,1;{\rm adj})}=\frac{2r}{r+1}$.
Several arguments imply that this perturbation corresponds to changing the
weight per unit length of the loops \cite{Fendley06}. This integrable
field theory describes the scaling limit of an integrable height model
\cite{DJMO}, and using the BWM algebra, it is described in
\cite{Fendley06} how to relate this height model to a dilute doubled
loop model very similar to the one we study here. Moving away from the
critical point in this similar model turns out to be effectively
changing the weight per unit length.
The second argument implying this result involves the $S$ matrices for
this integrable field theory, which decompose into the tensor product
of $S$ matrices of two minimal models ${\cal S}_{r} \times {\cal
S}_{r}$ \cite{Z91}. It is natural to interpret the worldlines of a
particle in a single minimal model as a loop in the $\On$ model
\cite{Z90}. Thus when the $S$ matrix is given by this tensor product,
it is natural to interpret the worldlines of such particles as doubled
loops; when two particles scatter they obey one of the four processes
in the vertices $w_1$, $w_2$ and $x$ pictured in Fig.~\ref{fig:model}.
In such an interpretation, the weight per unit length
of the loop is related to the mass of the particle. In the field
theory, moving along this integrable line corresponds precisely to
varying the mass of the particle.

\begin{figure}[th]
  \begin{center}
    \includegraphics{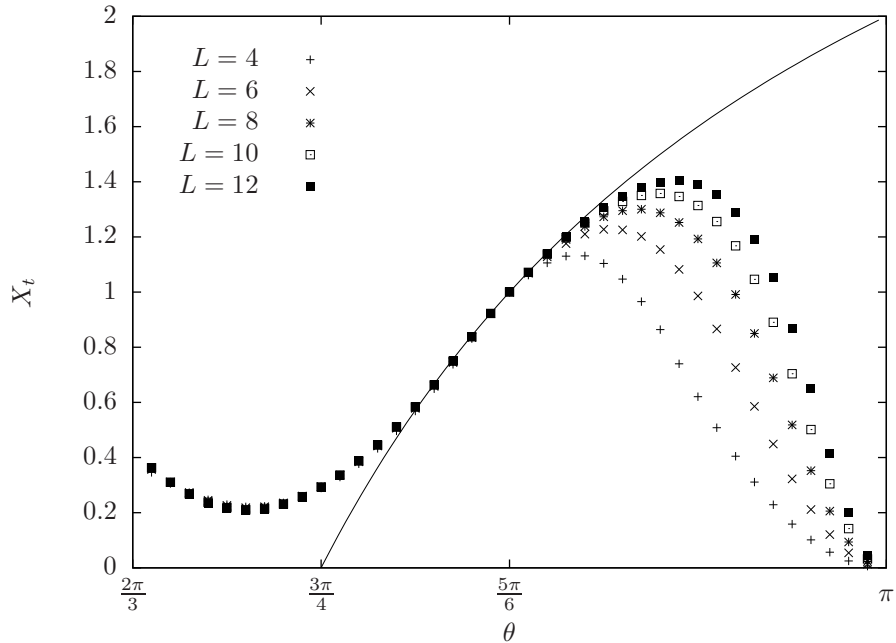}
    \caption{Thermal exponent $X_t$ in regime 4. Data points were obtained
      by transfer-matrix diagonalisation, and the solid line represents the
      exact result for the integrable perturbation dimension $X_{(1,1;{\rm adj})}=\frac{2r}{r+1}$.}
    \label{fig:xt}
  \end{center}
\end{figure}

We denote $X_t$ the thermal exponent, defined as the conformal dimension
for the first excited state in the zero-leg sector.
The numerical calculation of $X_t$ (see Fig.~\ref{fig:xt}) brings two observations.
In the region $\frac{5\pi}{6} \lesssim \theta < \pi$, the thermal exponent $X_t$ converges
to $X_{(1,1;{\rm adj})}=\frac{2r}{r+1}$ even for generic values of $\theta$, whereas $X_{(1,1;{\rm adj})}$
was derived only for integer values of $r$. This indicates that the results
from the $\SU(2)_r \times \SU(2)_r / \SU(2)_{2r}$ coset WZW model may be continued to arbitrary
$\frac{5\pi}{6} \lesssim \theta < \pi$. However, in the region $\frac{2\pi}{3} < \theta \lesssim \frac{5\pi}{6}$,
$X_t$ clearly deviates from the continued value $\frac{2r}{r+1}$: this shows that not
all exponents of the loop model are given by analytic continuation of the WZW
coset model in this region, including our point of interest $\theta=\frac{3\pi}{4}$.

\subsection{Correlation length exponent $\nu$}
\label{sec:nu}

The correlation length exponent $\nu$ is defined as the analog for the loop model of
$\nu_{\rm CC}$~\eqref{eq:nu-CC}. In the CC model, the effect of perturbing
the energy level $E$ away from the transition value $E_c$ amounts to taking
$\beta,\beta'$ out of the critical line~\eqref{eq:critical-beta}. The analog of this
perturbation in the integrable loop model is to introduce a staggering of the
spectral parameter between the even and odd sublattices, with symmetric values around
the isotropic spectral parameter $\varphi_0$~\eqref{eq:vf}:
$\varphi = (1 \pm \lambda)\varphi_0$, in the range $-1 \leq \lambda \leq 1$.

The parameter $\lambda$ acts in a similar way to $(\beta-\beta_c)$ in the original CC model.
At $\lambda=-1$, the only allowed loops are those with minimal length, winding around the
vertices of one sublattice (say, the even one). At $\lambda=1$ loops also have minimal length,
but wind around the odd sublattice. The critical transition takes place at $\lambda=0$, where
the two sublattices become equivalent, and loops may be very long.
In the limit $\lambda \to 0$, we expect this perturbation to develop a correlation length,
scaling as
\begin{equation} \label{eq:nu}
  \xi \sim |\lambda|^{-\nu} \,.
\end{equation}
Since this staggering does not
respect the rapidity lines of the square lattice, it breaks integrability.

Let us first discuss the point $\theta=\frac{5\pi}{6}$, where the model maps to free fermions
and remains solvable when the $\lambda$ perturbation is included (see Appendix~C).
At this point, we get the analytical result $\nu=2$, whereas the energy operator of
the free-fermion theory is $X_t=1$. In Appendix~C, we show that the effective
theory is a massive Majorana fermion with mass proportional to $\lambda^2$
and not $\lambda$. Thus, at $\theta=\frac{5\pi}{6}$, we have the relation
between $\nu$ and the dimension of the perturbing operator:
\begin{equation} \label{eq:X-nu}
  X_t = 2 - \frac{2}{\nu} \,.
\end{equation}
It is natural to assume that both $\nu$ and $X_t$ are continuous in $\theta$,
so the effective mass term should still be proportional to $\lambda^2$
outside $\theta=\frac{5\pi}{6}$, and the relation~\eqref{eq:X-nu} holds all along regime~4.

\begin{figure}[th]
  \begin{center}
    \includegraphics{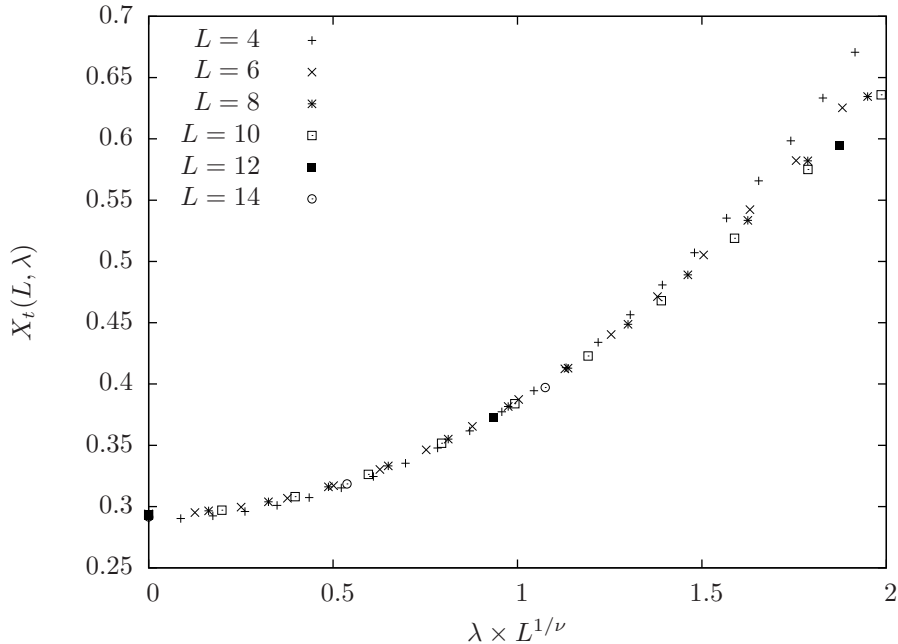}
    \caption{Data collapse for the effective thermal exponent $X_t(L,\lambda)$
      in the presence of a $\lambda$ perturbation. The value used for this plot is $1/\nu=0.9$.}
    \label{fig:nu}
  \end{center}
\end{figure}

For $\theta \neq \frac{5\pi}{6}$, exponent $\nu$ is only
accessible numerically, through finite-size scaling.
The correlation-length exponent $\nu$
is obtained by assuming a one-parameter scaling
law for the energy gap in the presence of the staggered perturbation
$\lambda$. For $\lambda \simeq 0$, we expect the behaviour:
\begin{equation}
  \log \frac{\Lambda_0(\lambda)}{\Lambda_1(\lambda)}
  \simeq \frac{2\pi}{L} F\left( \lambda \ L^{1/\nu} \right)\,,
\end{equation}
where $F$ is a scaling function.
Since eigenvalues are unchanged under $\lambda \to -\lambda$,
$F$ must be an even function. 
In particular, at $\theta=\frac{3\pi}{4}$, like for the truncated CC model
(see Sec.~\ref{sec:critical-prop}),
we get the best data collapse (see Fig.~\ref{fig:nu}) for the value
\begin{equation}
  \nu \simeq 1.1 \,.
\end{equation}
To our numerical precision, this value may be related through~\eqref{eq:X-nu} to
the thermal exponent $X_t \simeq 1.71$. This is an indication that the
relation~\eqref{eq:X-nu} should hold throughout regime 4.

\subsection{Other exponents}
\label{sec:other-exp}

\begin{figure}[th]
  \begin{center}
    \includegraphics{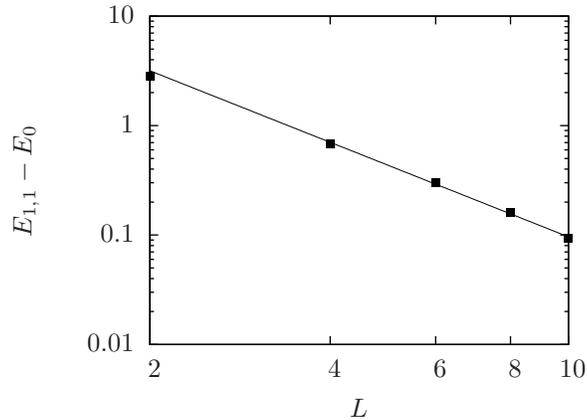}
    \caption{Log-log plot of the energy gap for the watermelon $(1,1)$ sector. Data points are
      fitted by a line of slope $\simeq -2.2$, and hence the conformal dimension is $X_{1,1}=0$.}
    \label{fig:x1}
  \end{center}
\end{figure}

A finite-size scaling plot of the gap corresponding to
$\ell_1=\ell_2=1$ is shown in Fig.~\ref{fig:x1}. The fact that it
scales to zero faster than $1/L$ indicates that $X_{1,1}$ is
consistent with zero. This agrees with the value $X_{|G|^2}=0$ expected
for the untruncated model in (\ref{eq:G2scaling}), but as far
as we know there is no fundamental reason for them to agree, and
it would be interesting to investigate this further. The observed
slope in Fig.~\ref{fig:x1} suggests existence of an irrelevant
operator with scaling dimension $\approx 3.2$.

Moreover, we observe numerically that $X_t=X_{2,2}$. Hence, like for usual dilute polymers,
this means that $X_t$ is associated to a perturbation of the monomer fugacity
(but different from the coset-model continuation to $r=0$, which would yield
$X_{(1,1;{\rm adj})}=0$), and the fractal dimension of a path is
\begin{equation}
  d_f = 2-X_{2,2} \simeq 1.71 \,.
\end{equation}

\section{Discussion}
\label{sec:discussion}

Regimes 3 and 4 of the integrable model are particularly
interesting, as the two loop colours remain coupled in the
continuum limit. They are described by the ``dilute branch'' of
the $\SU(2)_r \times \SU(2)_r / \SU(2)_{2r}$ WZW coset CFT (the same CFT as for the 
completely packed case~\cite{Paul-Jesper}), with an additional
Ising degree of freedom in regime 3. 
Strictly speaking, this theory is only valid at
the RSOS points, but some critical properties (including the
central charge) extend to the loop model for generic fugacity $n$.
However, differences between the loop and RSOS spectra exist, as
shown by our results on the thermal exponent $X_t$. An analytic
study of the loop model through the Bethe Ansatz Equations is
considered for future work. One also needs to understand if a
Coulomb-Gas construction (most probably with a three-dimensional
target space) could reproduce the coset results for generic $n$.

The original motivation for the present work was to propose an
exactly solvable approximation to the IQHE transition.
Unfortunately, our results for the correlation-length exponent
$\nu$ clearly indicate that the point $n=0$ of the integrable
model is not in the universality class of IQHE. However, this
model is a critical, integrable point in the phase diagram of our
modified CC model. It should really be considered as the first
order in a hierarchy of truncated models, converging to the IQHE
universality class. Higher-order truncated models should also
contain integrable points, which may be built by ``fusing'' the
edges of the first-order model, following~\cite{DJMO} or a more
recent approach based on additional $\mathbb{Z}_N$
symmetry~\cite{Z2}.

From the numerical point of view, we used the only known efficient method to study a generic
loop model: transfer-matrix diagonalisation. However, the inherent limitations on the system size
prevent us from obtaining sharp estimates for the exponents, especially for $\nu$. Recently,
new Monte-Carlo algorithms have been proposed to simulate 2d loop models~\cite{On-MC}.
We hope to adapt this new approach to two-colour models, and get more precise
estimates for the exponents of our truncated CC model.

\bigskip{\bf Acknowledgments}
The work of P.F. was supported in part by NSF grants
DMR/MPS 1006549 and 0704666.

\section*{Appendix A: The BWM algebra}
\renewcommand\thesection{A}
\setcounter{equation}{0}
\label{sec:BWM}

In this Appendix, we recall the motivation and definition of the BWM algebra in a graphical
language.

The BWM algebra~\cite{BWM} is a braid-monoid algebra, an object
relevant to knot theory. It was originally designed to compute a certain {\it link invariant},
and later it was realised that it could be represented by RSOS models related to affine
Lie algebras \cite{Wadati}. In~\cite{GW}, a dilute version of the BWM algebra was
constructed, together with the corresponding $R$-matrix.

\begin{figure}[th]
  \begin{center}
    \includegraphics{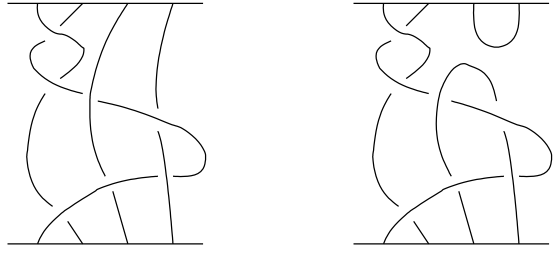}
    \caption{An element of the braid group (left) and a word of the braid-monoid algebra (right)
      for $L=4$.}
    \label{fig:braid}
  \end{center}
\end{figure}

In the context of knot theory, the basic objects under consideration are
{\it braids}. Let $(p_1, \dots, p_L)$ be $L$ distinct points in the
complex plane, and define two copies of each point in three-dimensional space,
$p'_j= p_j \times \{0\}, p''_j= p_j \times \{1\}$,
so that the points $\{p'_j\}$ and $\{p''_j\}$ lie in two parallel planes.
For all $j=1, \dots, L$, take a curve $\Gamma_j$ enclosed
between the two planes, and connecting $p'_j$ to $p''_j$.
Furthermore, impose that the $\Gamma_j$'s do not intersect each other.
Denote the multiplet $\Gamma = (\Gamma_1, \dots, \Gamma_L)$: a braid $\beta$ is then
an equivalence class of $\Gamma$'s, {\it modulo} continuous deformations of the curves.
A typical braid is depicted in Fig.~\ref{fig:braid}.

Multiplication of two braids $\beta, \beta'$
is defined by the concatenation of the two corresponding diagrams, with the convention
that diagrams act from bottom to top: the product $\beta \beta'$ corresponds
to $\beta$ above $\beta'$. The $\beta$'s form the braid group, generated by the
elementary braids $B_j, B_j^{-1}$, which satisfy the relations:
\begin{eqnarray}
  B_j B_j^{-1} &=& B_j^{-1} B_j = \idop \label{eq:braid1} \\
  B_j B_{j+1} B_j &=& B_{j+1} B_j B_{j+1} \label{eq:braid2} \\
  B_j B_\ell &=& B_\ell B_j \qquad \text{if $|j-\ell| \geq 2$.} \label{eq:braid3}
\end{eqnarray}

\begin{figure}[th]
  \begin{center}
    \includegraphics{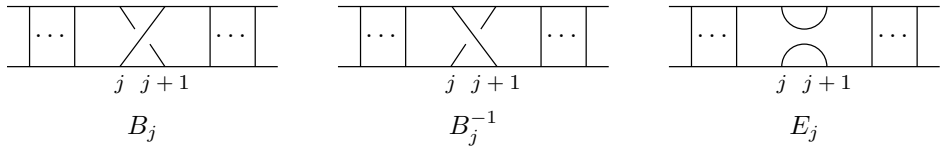}
    \caption{Generators of a braid-monoid algebra.}
    \label{fig:bm-gen}
  \end{center}
\end{figure}

\begin{figure}[th]
  \begin{center}
    \includegraphics{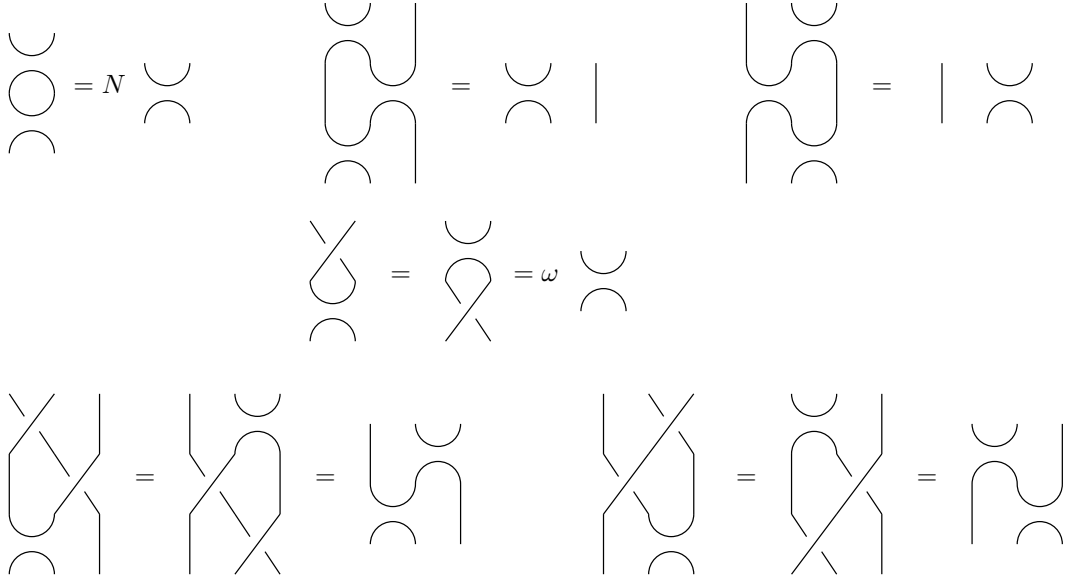}
    \caption{Algebraic rules in the braid-monoid algebra with parameters $(N,\omega)$.}
    \label{fig:bm-rules}
  \end{center}
\end{figure}

Consider now multiplets $\Gamma = (\Gamma_1, \dots, \Gamma_L)$
of non-intersecting curves connecting all the elements of $\{p'_j\} \cup \{p''_j\}$,
without the restriction that a curve should go from a $p'_j$ to a $p''_k$.
The corresponding diagrams are then
words on the alphabet $\{B_j, B^{-1}_j, E_j, \ j=1, \dots, (L-1)\}$, where the meaning
of the letters $B_j, B^{-1}_j, E_j$ is given in Fig.~\ref{fig:bm-gen}.
The algebra on these words is called a braid-monoid algebra.
It has two parameters $(N, \omega)$, and is defined by the braid-group
relations~\eqref{eq:braid1}--\eqref{eq:braid2}, together with the additional
relations (see Fig.~\ref{fig:bm-rules}):
\begin{eqnarray}
  E_j^2 &=& N \ E_j \label{eq:TL1} \\
  E_j E_{j \pm 1} E_j &=& E_j \label{eq:TL2} \\
  B_j E_j &=& E_j B_j = \omega \ E_j \label{eq:bm1} \\
  B_j B_{j \pm 1} E_j &=& E_{j \pm 1} B_j B_{j\pm 1} = E_{j \pm 1} E_j \label{eq:bm2} \,.
\end{eqnarray}
Equations~\eqref{eq:TL1}--\eqref{eq:TL2} mean that the $E_j$ form a Temperley-Lieb algebra
with loop weight $N$.

The BWM algebra is a braid-monoid algebra \eqref{eq:braid1}--\eqref{eq:bm2}
where one imposes a linear relation between $B_j$, $B_j^{-1}$ and $E_j$:
\begin{equation} \label{eq:BWM}
  E_j = \idop + \frac{B_j - B_j^{-1}}{q-q^{-1}} \,,
  \qquad \text{where}
  \quad N = 1+ \frac{\omega-\omega^{-1}}{q-q^{-1}} \,.
\end{equation}
The reason for introducing such a constraint is that the resulting algebra
supports a linear form (the Markov trace) which is identical to a geometric invariant
of the diagrams $\Gamma$ (the Kauffman polynomial)~\cite{BWM}.

\begin{figure}[th]
  \begin{center}
    \includegraphics{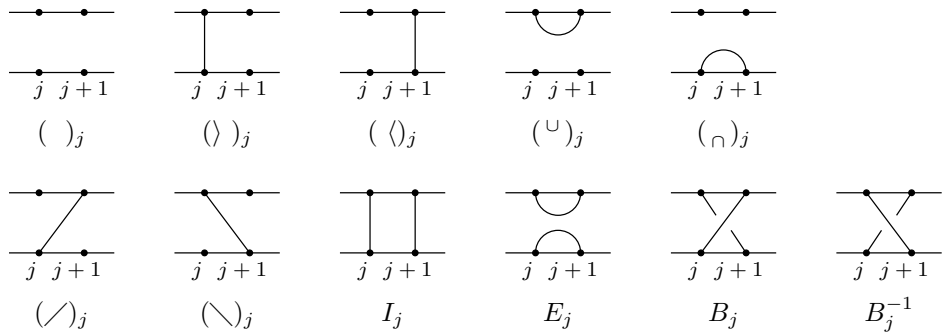}
    \caption{The generators of the dilute BWM algebra.}
    \label{fig:dBWM-gen}
  \end{center}
\end{figure}

The dBWM algebra~\cite{GW} is obtained by allowing vacancies, or equivalently by taking
multiplets of curves $\Gamma=(\Gamma_1, \dots, \Gamma_\ell)$ with $0 \leq \ell \leq L$.
This amounts to adding the generators
\begin{equation*}
  I_j, \vac_j, (\rangle \ )_j, (\ \langle )_j,
  (\begin{smallmatrix} \cup \\ \phantom{\cap} \end{smallmatrix})_j,
  (\begin{smallmatrix} \phantom{\cup} \\ \cap \end{smallmatrix})_j,
  (\diagup)_j, (\diagdown)_j,
\end{equation*}
whose action is depicted in Fig.~\ref{fig:dBWM-gen}. In equations~\eqref{eq:braid1}
and \eqref{eq:BWM},
$\idop$ is replaced by $I_j$, so that the $B_j, B^{-1}_j, E_j$ still form a BWM algebra
on the set of occupied sites. Additional relations for the dilute generators should be
included, to implement invariance under continuous deformation of the curves in the
presence of vacancies. The full set of dBWM relations is given in~\cite{GW}.

\section*{Appendix B: Discretely holomorphic parafermion in the loop model}
\renewcommand\thesection{B}
\setcounter{equation}{0}
\label{sec:holo}

In this Appendix, we show that the two-colour loop model admits a
discretely holomorphic para\-fermion $\psi_s(z)$~\cite{Riva,Raj,IC} exactly on
the integrable manifold~\eqref{eq:weights}.
The parafermion $\psi_s(z)$ is defined on the midpoints of the dual lattice ${\cal L}^*$,
and inserts a one-leg defect for each color at point $z$.
In the two-point function $\langle \psi_s(0) \psi_s(z) \rangle$, there is a black
({\resp} grey) path $\gamma_1$ ({\resp} $\gamma_2$) connecting $0$ and $z$, and one
includes a phase factor involving the winding angles $W$ of the paths $\gamma_1, \gamma_2$:
\begin{equation}
  \langle \psi_s(0) \psi_s(z) \rangle = \frac{1}{Z} \sum_{(\gamma_1,\gamma_2)_{0 \to z}}
  \ \sum_{C|(\gamma_1,\gamma_2)} \Pi(C) \ e^{\frac{is}{2} [W(\gamma_1)+W(\gamma_2)]} \,,
\end{equation}
where the first sum is over all possible pairs of paths from $0$ to $z$, the
second sum is over the loop configurations $C$ compatible with $\gamma_1$ and $\gamma_2$,
and $\Pi(C)$ is the Boltzmann weight for a loop configuration $C$.

\begin{figure}[th]
  \begin{center}
    \includegraphics{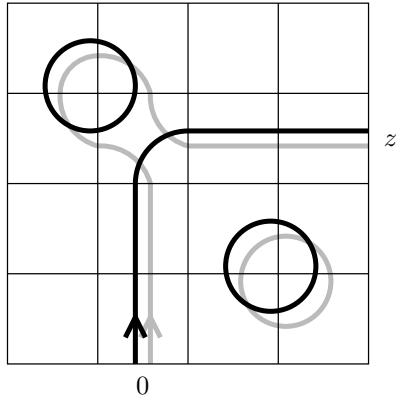}
    \caption{A loop configuration contributing to $\langle \psi_s(0) \psi_s(z)\rangle$.}
    \label{fig:parafermion}
  \end{center}
\end{figure}

We impose discrete Cauchy-Riemann (CR) on $\langle \psi_s(0) \psi_s(z) \rangle$:
\begin{equation} \label{eq:CR}
  \sum_{z \in \square} \langle \psi_s(0) \psi_s(z) \rangle \ \delta z = 0 \,,
\end{equation}
where the sum is over the edges of an elementary plaquette of ${\cal L}^*$, and the
$\delta z$'s are the corresponding elementary displacements.
For the discrete CR equations~\eqref{eq:CR} to hold, it is sufficient to fix the external
loop configuration outside a given plaquette, and ask the total contribution
of internal configurations to vanish~\cite{Riva,IC}.
This determines a linear system of equations for the Boltzmann weights.
To get anisotropic solutions, we consider the analog problem on a rhombic
lattice of angle $\alpha$ \cite{Raj,IC}.
Setting
\begin{equation}
  \lambda:= e^{\frac{i \pi s}{2}} \,, \qquad \mu:= e^{i\alpha(1+s)}\,,
\end{equation}
we get the $7 \times 7$ linear system for the unknowns $(t,u_1,u_2,v,w_1,w_2,x)$:
\begin{equation}
  \label{eq:holo}
\begin{array}{rcl}
  t+ \mu \lambda^{-2} u_1 - \mu u_2 - v &=& 0 \\
  n^2 u_1 - \lambda^{-2} u_2 - \mu \lambda^2 v
  + \mu \lambda^{-2} (n^2 w_1 + w_2 + 2nx) &=& 0 \\
  -\lambda^2 u_1 + n^2 u_2 + \mu \lambda^{-4} v
  - \mu (w_1 + n^2 w_2 + 2nx) &=& 0 \\
  -\mu \lambda^2 u_1 + \mu \lambda^{-4} u_2 + n^2 v
  - \lambda^4 w_1 - \lambda^{-4} w_2 - 2x &=& 0 \\
  n \lambda u_1 - n \lambda^{-1} u_2 - \mu \lambda^{-1} v + \mu \lambda^{-1}
  [n(w_1+w_2) + (n^2+1)x] &=& 0 \\
  n \lambda^{-1} u_1 - \lambda u_2 - n \mu \lambda v + \mu \lambda^{-3} (n w_1+x)
  + \mu \lambda (w_2+nx) &=& 0 \\
  -u_1 + n \lambda^2 u_2 + n \mu \lambda^{-2} v - \mu \lambda^{-2} (w_1+nx)
  - \mu \lambda^2 (n w_2+x) &=& 0 \,,
\end{array}
\end{equation}
As a first step, we need determine the spin $s$ by going back to the isotropic case
$\alpha=\pi/2$. Imposing $u_1=u_2$ and $w_1=w_2$, \eqref{eq:holo} reduces to a $5 \times 5$
system, whose determinant is:
\begin{equation} \label{eq:det}
  D(n, \lambda) = -\lambda^{-4} (\lambda^2+1)(n\lambda^4-1)^2
  (\lambda^4+\lambda^{-4} + n^3-3n) \,.
\end{equation}
Using the parameterisation $n= -2 \cos 2\theta$, this determinant vanishes when:
\begin{equation} \label{eq:s}
  \exp(2i\pi s) = \exp(\pm 6i\theta) \,.
\end{equation}

For a general angle $\alpha$, the solution of~\eqref{eq:holo} is a set of $\alpha$-dependent
weights $t(\alpha), \dots x(\alpha)$. If we apply the substitution:
\begin{equation} \label{eq:alpha-phi}
  \alpha \to \frac{\varphi}{1+s} \,,
\end{equation}
we observe that the solution of~\eqref{eq:holo} is identical to the integrable
weights~\eqref{eq:weights}. This is analogous to what was found for various
other integrable models with a discrete holomorphic parafermion~\cite{Riva,Raj,IC}.
Note that the relation~\eqref{eq:alpha-phi} is consistent
with the discussion on the Fermi velocity in Section~\ref{sec:Hamiltonian}.

Moreover, at the Ising points $\theta=\frac{\pi}{6}, \frac{5\pi}{6}$ (see Appendix~C), the spin $s=\frac{1}{2}$
is consistent with~\eqref{eq:s}.

\begin{figure}[th]
  \begin{center}
    \includegraphics[scale=0.8]{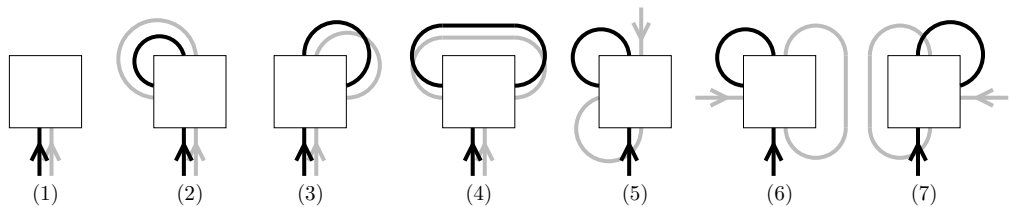}
    \caption{External loop connectivities outside an elementary plaquette.}
    \label{fig:holo}
  \end{center}
\end{figure}

\section*{Appendix C: Free fermions at $\theta= \frac{\pi}{6}, \frac{5\pi}{6}$}
\renewcommand\thesection{C}
\setcounter{equation}{0}
\label{sec:free-fermions}

\subsection{Mapping to a staggered 8V model}

\begin{figure}[th]
  \begin{center}
    \includegraphics[scale=1]{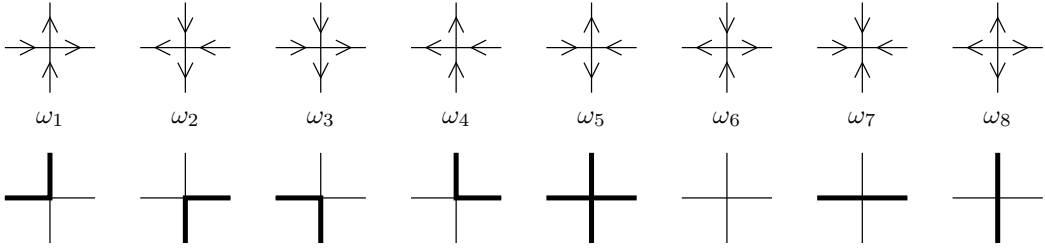}
    \caption{Correspondence between the 8V model and the ${\rm O}(n=1)$ model.
      The mapping depicted here is valid on one sublattice, say even sites.
      On odd sites, all arrows must be reversed.}
    \label{fig:8V}
  \end{center}
\end{figure}

At $\theta=\frac{\pi}{6}$ and $\theta=\frac{5\pi}{6}$, the model maps to a free fermion
Hamiltonian through the mapping sequence:
\begin{center}
  two-color loop $\to$ square-lattice ${\rm O}(n=1)$ $\to$ staggered 8V $\to$ free fermions.
\end{center}
In both cases, we start from the two-color loop model and do the sign change
$(n,x) \to (-n,-x)$ to get $n=1$. With this value of $n$, one may discard loop connectivities,
and the model is simply local, with occupied and empty edges. After the exchange of
occupied/empty edges, we get a square-lattice $\On$ model~\cite{square-On} with $n=1$
and weights~:
\begin{equation}
  \wt{t} = w_1+w_2 -2x,
  \quad \wt{u}_1=u_1,
  \quad \wt{u}_2=u_2,
  \quad \wt{v}=v,
  \quad \wt{w}_1+\wt{w}_2= t \,.
\end{equation}
(The tilde is here to avoid confusion between the $\On$ and two-colour
loop model Boltzmann weights).
For $\theta= \frac{\pi}{6}$, we have the specific values
\begin{equation}
  \begin{array}{rcl}
    \wt{t}_{\phantom{1}} &=& \sin 2\varphi -\sqrt{3} \\
    \wt{u}_1 &=& \phantom{-} \sqrt{3} \ \cos \varphi \\
    \wt{u}_2 &=& - \sqrt{3} \ \sin \varphi \\
    \wt{v}_{\phantom{1}} &=& \sin 2\varphi \\
    \wt{w}_1 &=& - \cos(\frac{\pi}{6}-2\varphi) - \frac{\sqrt{3}}{2} \\
    \wt{w}_2 &=& \phantom{-} \cos(\frac{\pi}{6}+2\varphi) - \frac{\sqrt{3}}{2} \,,
  \end{array}
\end{equation}
which are exactly those of the integrable square-lattice $\On$ model~\cite{square-On}
with $n=1$ at the dilute critical point. The ${\rm O}(n=1)$ model maps in turn to an
eight-vertex model (see Fig.~\ref{fig:8V}) with staggered weights
\begin{equation} \label{eq:8V}
  \begin{array}{rcl}
    \omega_1=\omega_2 &=& \phantom{-}\sqrt{3} \cos\varphi \\
    \omega_3=\omega_4 &=& -\sqrt{3} \sin\varphi \\
    (\omega_5,\omega_6) &=& \begin{cases}
      (-\sin 2\varphi -\sqrt{3}, \phantom{-}\sin 2\varphi -\sqrt{3}) & \hbox{on even sites} \\
      (\phantom{-}\sin 2\varphi -\sqrt{3}, -\sin 2\varphi -\sqrt{3}) & \hbox{on odd sites}
    \end{cases} \\
    \omega_7=\omega_8 &=& \sin 2\varphi \,.
  \end{array}
\end{equation}
At $\theta= \frac{5\pi}{6}$, one gets the same 8V model, up to irrelevant signs.

\subsection{Very anisotropic limit: the XY chain in a magnetic field}

In terms of the Pauli matrices $\sigma_j$, the 8V $\Rc$-matrix reads:
\begin{eqnarray}
  \Rc_j^{\rm 8V} &=& \frac{1}{4}(\omega_1+\omega_2+\omega_5+\omega_6) \idop
  + \frac{1}{4}(\omega_1+\omega_2-\omega_5-\omega_6) \sigma_j^z \sigma_{j+1}^z \nonumber \\
  &&+ (\omega_3 \ \sigma_j^- \sigma_{j+1}^+ + \omega_4 \ \sigma_j^+ \sigma_{j+1}^-
  + \omega_7 \ \sigma_j^- \sigma_{j+1}^- + \omega_8 \ \sigma_j^+ \sigma_{j+1}^+) \nonumber \\
  &&+ \frac{1}{4}(\omega_1-\omega_2) (\sigma_j^z+\sigma_{j+1}^z)
  + \frac{1}{4}(\omega_5-\omega_6) (\sigma_j^z-\sigma_{j+1}^z) \,.
\end{eqnarray}
We can now take the very anisotropic limit $\varphi \to 0$. Denoting by a prime the derivative
with respect to $\varphi$ at $\varphi=0$, and using the weights~\eqref{eq:8V}, we obtain:
\begin{equation}
  \Rc_j^{'\rm 8V} = -\sqrt{3} (\sigma_j^- \sigma_{j+1}^+ + \sigma_j^+ \sigma_{j+1}^-)
  + 2(\sigma_j^- \sigma_{j+1}^- + \sigma_j^+ \sigma_{j+1}^+)
  - (-1)^j (\sigma_j^z-\sigma_{j+1}^z) \,.
\end{equation}
The critical Hamiltonian is given by
\begin{equation} \label{eq:H0}
  H_0 = -\half \sum_{j=1}^L \Rc_j^{'\rm 8V}
  = - \frac{1}{4} \sum_{j=1}^L \left[
    J_x \ \sigma_j^x \sigma_{j+1}^x - J_y \ \sigma_j^y \sigma_{j+1}^y
    + 2 h \ (-1)^j \sigma_j^z
  \right] \,,
\end{equation}
where $J_x= (2-\sqrt{3})$, $J_y= (2+\sqrt{3})$ and $h=2$.
The alternating sign of the last term in~\eqref{eq:H0} can be eliminated
by the unitary change of basis:
\begin{equation}
  H_0 \to U^\dag H_0 U \,,
  \qquad \text{where} \qquad
  U := \prod_{\ell=1}^{L/2} \sigma^x_{2\ell-1} \,.
\end{equation}
This maps $H_0$ to an XY chain in a magnetic field~\cite{XY,Ising}:
\begin{equation} \label{eq:H-XY}
  H_{\rm XY} = U^\dag H_0 U = -\half \sum_{j=1}^L \left[
    (1+\gamma) \ \sigma_j^x \sigma_{j+1}^x + (1-\gamma) \ \sigma_j^y \sigma_{j+1}^y
    + h \ \sigma_j^z
  \right] \,,
\end{equation}
where $\gamma=-\frac{\sqrt{3}}{2}$ and $h=2$. In particular, we have learnt that this particular point
of the XY spin chain is exactly equivalent to the integrable dilute ${\rm O}(n=1)$ model.

\subsection{Staggered perturbation associated to $\nu$}

The setting of the 8V model also allows us to consider a staggered perturbation
like the one defining
exponent $\nu$ (see Sec.~\ref{sec:nu}). To do this, we introduce staggered
spectral parameters $(1 \pm \lambda) \varphi$ on the even/odd sites. We take the very
anisotropic limit $\varphi \to 0$,
with $\lambda$ fixed. This way, the parameter $\lambda$ controls the strength of
the perturbation.
The resulting Hamiltonian has the form
$$H(\lambda) = H_0 + \lambda H_1 \,$$
where
\begin{equation} \label{eq:H1}
  H_1 = -\half \sum_{j=1}^L (-1)^j \Rc_j^{'\rm 8V}
  = -\frac{1}{4} \sum_{j=1}^L (-1)^j \ \left(
    J_x \ \sigma_j^x \sigma_{j+1}^x - J_y \ \sigma_j^y \sigma_{j+1}^y
  \right) \,,
\end{equation}
and $J_x,J_y$ are the same as for $H_0$.
After the unitary change of basis defined by $U$, we get the
perturbing term:
\begin{equation} \label{eq:Hp}
  H_p = U^\dag H_1 U =
  - \frac{\lambda}{2} \sum_{j=1}^{L} 
    (-1)^j
    \left[ (1+\gamma) \ \sigma_j^x \sigma_{j+1}^x
      + (1-\gamma) \ \sigma_j^y \sigma_{j+1}^y
    \right] \,.
\end{equation}

\subsection{Exact free-fermion solution}

In this paragraph, we expose the exact solution of $H(\lambda)$ for general
values of $\gamma,h,\lambda$.
Following~\cite{XY}, we can solve the model $H(\lambda)$ by a Jordan-Wigner
transformation, mapping the Pauli matrices~$\sigma_j$ to fermion operators
\begin{equation}
  c_j := \left( \prod_{\ell=1}^{j-1} \sigma^z_\ell \right) \ \sigma^+_j \,,
  \qquad c^\dag_j := \left( \prod_{\ell=1}^{j-1} \sigma^z_\ell \right) \ \sigma^-_j \,,
  \qquad c^\dag_j c^{\phantom \dag}_j = \half(1-\sigma^z_j) \,,
\end{equation}
obeying anti-commutation relations:
\begin{equation}
  \{ c_j, c_\ell \}=0 \,,
  \qquad \{ c_j, c^\dag_\ell \} = \delta_{j \ell} \,.
\end{equation}
In this language, the perturbed Hamiltonian reads
\begin{equation}
  H(\lambda) = 
  - \sum_{j=1}^L \Bigg\{
  \left[ 1 + (-1)^j \lambda \right]
  \left[
    c^\dag_j c^\nodag_{j+1} + c^\dag_{j+1} c^\nodag_j
    + \gamma (c^\dag_j c^\dag_{j+1} + c_{j+1} c_j)
  \right]
  + h \left(\half - c^\dag_j c^\nodag_j \right)
  \Bigg\} \,.
\end{equation}
We introduce two species of fermions
\begin{equation}
  c_{1,\ell} := c_{2\ell} \,,
  \qquad
  c_{2,\ell} := c_{2\ell-1} \,,
\end{equation}
and their Fourier modes
\begin{equation}
  c_{\mu,q} := \frac{1}{\sqrt{L/2}}
  \sum_{\ell=1}^{L/2} e^{i\ell q} c_{\mu,\ell} \,,
  \qquad q = \frac{2\pi m}{L/2} - \pi \,,
  \quad m=1, \dots, \frac{L}{2} \,.
\end{equation}
We can now rewrite $H(\lambda)$ as
\begin{equation}
  H(\lambda) = \sum_{-\pi < q \leq \pi}
  \sum_{\mu,\nu} \left[
    c^\dag_{\mu,q} A^\nodag_{\mu \nu, q} c^\nodag_{\nu,q}
    + \half \left( c^\dag_{\mu,q} B^\nodag_{\mu \nu, q} c^\dag_{\nu,-q}
    - c_{\mu,-q} B_{\mu \nu, q} c_{\nu,q} \right)
  \right] \,,
\end{equation}
where the matrices $A_q$ and $B_q$ read
\begin{equation}
  A_q = \left( \begin{array}{cc}
      h & \alpha_q^* \\
      \alpha_q & h
    \end{array} \right) \,, 
  \qquad B_q = \left( \begin{array}{cc}
      0 & -\beta^*_q \\
      \beta_q & 0
    \end{array} \right) \,,
  \qquad \begin{array}{rcl}
    \alpha_q &:=& -[(1+\lambda)e^{iq} + (1-\lambda)] \,, \\
    \beta_q &:=& \gamma[(1+\lambda)e^{iq} - (1-\lambda)] \,.
  \end{array}
\end{equation}
Like in~\cite{XY}, the energies $\epsilon_\mu(q)$ are the square roots of
the eigenvalues of $(A_q+B_q)(A_q-B_q)$:
\begin{equation} \label{eq:epsq}
  \epsilon_{1,2}(q) = 2 \sqrt{
    \frac{h^2}{4} + (1+ \gamma^2 \lambda^2) \cos^2 \frac{q}{2}
    + (\gamma^2 + \lambda^2) \sin^2 \frac{q}{2}
    \mp  \sqrt{
      4 \gamma^2 \lambda^2 + h^2
      \left| \cos \frac{q}{2} + i\lambda \sin^2 \frac{q}{2} \right|^2
    }
  } \,.
\end{equation}
This is the two-branch dispersion relation for arbitrary $\gamma,h,\lambda$.

To express the corresponding eigenmodes, we need the unitary $2 \times 2$
matrices $W_q,V_q$ defined by the linear relations
\begin{equation}
  (A_q - B_q) W_q = V_q D_q \,,
  \qquad (A_q + B_q) V_q = W_q D_q \,,
  \qquad D_q := \left(\begin{array}{cc}
      \epsilon_1(q) & 0 \\
      0 & \epsilon_2(q)
    \end{array} \right) \,.
\end{equation}
The Bogoliubov transformation diagonalising $H(\lambda)$ is
\begin{equation} \label{eq:eta}
  \eta_{\mu,q} := \half \sum_{\nu} \left[(W+V)^\dag_{\mu\nu,q} \, c^\nodag_{\nu,q}
    - (W-V)^\dag_{\mu\nu,q} \, c^\dag_{\nu,-q}
  \right] \,.
\end{equation}
Unitarity of $V_q$ and $W_q$ ensures the canonical
anticommutation relations
\begin{equation}
  \{ \eta_{\mu,q}, \eta_{\mu',q'} \} = 0 \,,
  \qquad \{ \eta^\nodag_{\mu,q}, \eta^\dag_{\mu',q'} \}
  = \delta_{\mu \mu'} \delta_{q q'} \,.
\end{equation}
In terms of the $\eta$'s, the Hamiltonian reads
\begin{equation}
H(\lambda) = \sum_{-\pi < q \leq \pi}
  \ \sum_{\mu=1,2} \epsilon_\mu(q)
  \, \eta^\dag_{\mu,q} \eta^\nodag_{\mu,q} \,.
\end{equation}
Note that there are $L/2$ distinct momenta $q$,
and that each momentum corresponds to two modes $\mu=1,2$. Thus,
we recover $L$ independent modes $\eta_{\mu,q}$.

As a final step, we perform the change
\begin{equation} \label{eq:etah}
  \eta_{\mu,q} \to \wt{\eta}_{\mu,q} = \begin{cases}
    \eta_{\mu,q} & \text{if $q \geq 0$,} \\
    \eta^\dag_{\mu,q} & \text{if $q < 0$,}
  \end{cases}
\end{equation}
so that the modes with $q<0$ are now considered as holes.
the Hamiltonian becomes
\begin{equation}
  H(\lambda) = \sum_{-\pi < q \leq \pi}
  \ \sum_{\mu=1,2} \wt{\epsilon}_\mu(q)
  \, \wt{\eta}^\dag_{\mu,q} \wt{\eta}^\nodag_{\mu,q} \,,
  \qquad \text{where}
  \qquad \wt{\epsilon}_\mu(q) := {\rm sgn}(q) \, \epsilon_\mu(q) \,.
\end{equation}

\subsection{Critical Majorana fermion at $h=2, \lambda=0$}

\begin{figure}[th]
  \begin{center}
    \begin{tabular}{cc}
      \includegraphics[scale=1]{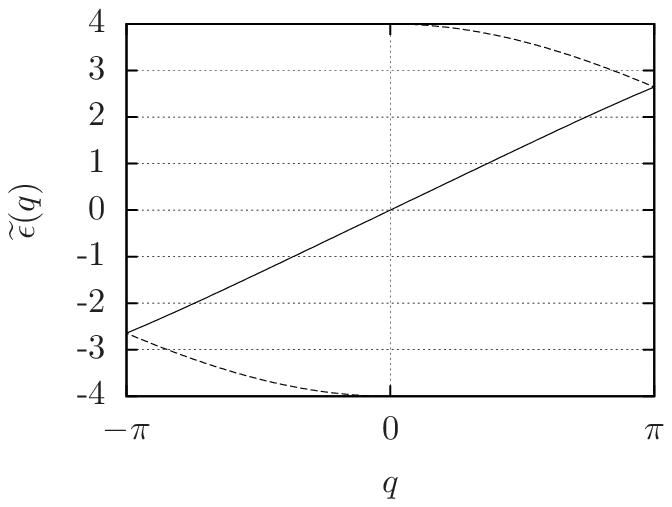}
      & \includegraphics[scale=1]{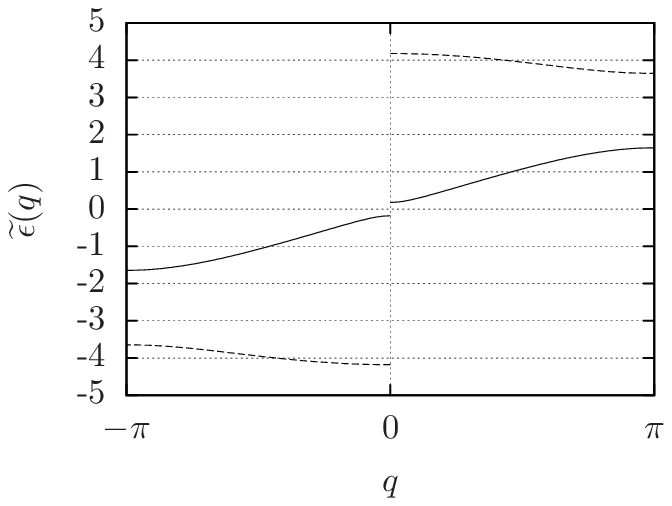}
    \end{tabular}
    \caption{The dispersion relation for $h=2, \gamma=-\frac{\sqrt{3}}{2}$.
      Left: critical case $\lambda=0$. Right: $\lambda=\half$. The full ({\resp} dotted) lines
    represent $\wt{\epsilon}_1$ ({\resp} $\wt{\epsilon}_2$).}
    \label{fig:eps}
  \end{center}
\end{figure}

The dispersion relation of the XY chain in
a magnetic field is obtained by setting $\lambda=0$ in~\eqref{eq:epsq}:
\begin{equation}
  \epsilon_{1,2}(q) = 2 \sqrt{ 
    \left(\frac{h}{2} \mp \cos \frac{q}{2} \right)^2
    + \gamma^2 \sin^2 \frac{q}{2}
  }
  \qquad \qquad (\lambda=0) \,.
\end{equation}
For $h=2$, $\epsilon_1$ is critical at $q=0$, whereas $\epsilon_2$ is not critical\footnote{
  In the case $h=\gamma=0$, both modes $\epsilon_1,\epsilon_2$ are critical at $q=\pi$,
  and the corresponding field theory is a critical Dirac fermion.
}:
\begin{eqnarray}
  \wt{\epsilon}_1(q) &=&
  4 \sin \frac{q}{4} \sqrt{\sin^2 \frac{q}{4}
    + \gamma^2 \cos^2 \frac{q}{4}} \\
  \wt{\epsilon}_2(q) &=&
  4 \, {\rm sgn}(q) \cos \frac{q}{4} \sqrt{\cos^2 \frac{q}{4}
    + \gamma^2 \sin^2 \frac{q}{4}} \,.
\end{eqnarray}
The dispersion relation $\wt{\epsilon}_1(q)$ is approximately linear at $q=0$
(see Fig.~\ref{fig:eps}).
In the ground state, all levels with $-\pi<q<0$ are filled: this is a
Fermi sea with only one Fermi level $q_f=0$, and thus it corresponds
to a Majorana fermion with central charge $c=\half$.
The critical eigenmodes are obtained from~\eqref{eq:eta}:
\begin{equation} \label{eq:etac}
  \left\{ \begin{array}{rcl}
    \eta_{1,q} &=& \cos \frac{\theta_{1,q}}{2} \, (c_{1,q} + c_{2,q})
    + i \sin \frac{\theta_{1,q}}{2} \, (c^\dag_{1,-q} + c^\dag_{2,-q}) \,, \\
    \eta_{2,q} &=& \cos \frac{\theta_{2,q}}{2} \, (c_{1,q} - c_{2,q})
    + i \sin \frac{\theta_{2,q}}{2} \, (c^\dag_{1,-q} - c^\dag_{2,-q}) \,,
  \end{array} \right.
\end{equation}
where
\begin{equation} \label{eq:thetaq}
  \left\{ \begin{array}{rcl}
      \theta_{1,q} &:=& {\rm Arg} \left(
        1-\cos\frac{q}{2} - i \sin \frac{q}{2} \right) \,, \\
      \theta_{2,q} &:=& {\rm Arg} \left(
        1+\cos\frac{q}{2} + i \sin \frac{q}{2} \right) \,.
    \end{array} \right.
\end{equation}
After the change $\eta \to \wt{\eta}$~\eqref{eq:etah}, the continuum limit is
described by the effective Hamiltonian
\begin{equation}
  H_0 \simeq i v_f \int {\rm d}x \ \wt{\eta}_1^\dag \partial_x \wt{\eta}^\nodag_1 \,.
\end{equation}

\subsection{Gapped theory at $h=2,\lambda>0$}

When the $\lambda$ perturbation is turned on, an energy gap opens at
$q=0$ (see Fig.~\ref{fig:eps}):
\begin{equation} \label{eq:DE}
  \Delta E := 2 \epsilon_1(q=0) = 2 \left|
    \sqrt{1+\gamma^2 \lambda^2} - 1
  \right| \sim \gamma^2 \lambda^\nu \,,
  \qquad \nu=2 \,.
\end{equation}
To understand this value of $\nu$, we shall analyse the perturbing term $H_p$ in
terms of the critical modes $\wt{\eta}_{1,q}$~\eqref{eq:etac}.
The relations~\eqref{eq:etac} can be inversed, to give:
\begin{equation} \label{eq:aq}
  \left\{\begin{array}{rcl}
    c_{1,q}+c_{2,q} &=& \cos \frac{\theta_{1,q}}{2} \, \eta_{1,q}
    - i \sin \frac{\theta_{1,q}}{2} \, \eta^\dag_{1,-q} := a_{1,q} \,, \\
    c_{1,q}-c_{2,q} &=& \cos \frac{\theta_{2,q}}{2} \, \eta_{2,q}
    - i \sin \frac{\theta_{2,q}}{2} \, \eta^\dag_{2,-q} := a_{2,q} \,.
  \end{array} \right.
\end{equation}
The perturbing term has the expression
\begin{eqnarray}
  H_p = \half \sum_{-\pi< q \leq \pi} \Bigg\{ &&
  (1-\cos q) \, (a_{1,q}^\dag a^\nodag_{1,q} - a_{2,q}^\dag a^\nodag_{2,q}) \nonumber \\
  && + i\sin q \, \left[
    a_{2,q}^\dag a^\nodag_{1,q} +
    \frac{\gamma}{2} \, (a_{1,q}^\dag a_{1,-q}^\dag - a_{2,q}^\dag a_{2,-q}^\dag)
    - {\rm h.c.}
  \right] \nonumber \\
  && -\gamma (1+\cos q) \, (a_{2,q}^\dag a_{1,-q}^\dag + a_{1,-q} a_{2,q})
  \quad \Bigg\} \,.
  \label{eq:Hp2}
\end{eqnarray}
In the region $q \simeq 0$, the first term in~\eqref{eq:Hp2} is of order $q^2$,
and thus it generates irrelevant terms of the form $\wt{\eta}_1^\dag \partial_x^2 \wt{\eta}_1^\nodag$ in
the continuum limit. The second term is of order $q$, and corresponds to
$\wt{\eta}_1^\dag \partial_x \wt{\eta}_1^\nodag$, which renormalises the Fermi velocity.
At first order in $\lambda$, the third term has no effect on the continuum theory.
However, in second-order perturbation in $\lambda$, it generates terms of
the form $(\wt{\eta}^\dag_2 \wt{\eta}^\nodag_2)(\wt{\eta}^\dag_1 \wt{\eta}^\nodag_1)$, which are non-vanishing
since the lowest $\wt{\eta}_2$ modes are occupied in the ground state.
From this analysis, we obtain the effective Hamiltonian in the continuum limit
\begin{equation}
  H_{\rm eff}(\lambda) \propto
  \int {\rm d}x \ \left(i \, \wt{\eta}_1^\dag \partial_x \wt{\eta}^\nodag_1
    + {\rm const} \times \gamma^2 \lambda^2
    \, \wt{\eta}_1^\dag \wt{\eta}^\nodag_1 \right) \,.
\end{equation}
The ``mass term'' has dimension $X_t=1$, and the energy gap thus scales as
$$
\Delta E \propto \left(\lambda^2 \right)^\frac{1}{2-X_t} \,.
$$
Comparing with~\eqref{eq:DE}, we get the scaling relation
\begin{equation}
  X_t = 2 - \frac{2}{\nu} \,.
\end{equation}

\end{document}